\documentclass[useAMS,usenatbib]{mn2e}

\usepackage{graphicx}
\usepackage{times}
\usepackage{footnote}
\usepackage{microtype}
\usepackage{multirow}
\usepackage{multicol}

\title[Periodic Spectroscopic Variability of FU Ori]{The Periodic Spectroscopic Variability of FU Orionis\thanks{Based on observations made at Observatoire de Haute-Provence (CNRS), France}}
\author[S. L. Powell et al.]{Stacie L. Powell$^{1}$\thanks{E-mail: slp65@ast.cam.ac.uk}, 
Mike Irwin${^1}$, Jerome Bouvier${^2}$ and Cathie J. Clarke${^1}$\\
$^{1}$Institute of Astromomy, Madingley Road, Cambridge CB3 OHA, UK\\
$^{2}$ UJF-Grenoble 1 / CNRS-INSU, Institut de Plan\'{e}tologie et d’Astrophysique de Grenoble (IPAG) UMR 5274, Grenoble, F-38041, France}

\begin{document}


\pagerange{\pageref{firstpage}--\pageref{lastpage}} \pubyear{2012}

\maketitle

\label{firstpage}

\begin{abstract}
FU Orionis systems are young stars undergoing outbursts
of disc accretion and where the optical spectrum contains
lines associated with both the disc 
photosphere and a wind
component. Previous observations of the prototype FU Orionis
have suggested that the wind lines and the photospheric
lines are 
modulated with periods of $14.54$ and $3.54$ days
respectively \citep{herbig03}. We have re-observed the
system at higher spectral resolution, by 
monitoring variations
of optical line profiles over $21$ nights in 2007 and have
found  periods of $13.48$ and $3.6$ days in the wind
and disc components, 
consistent with the above: this implies
variability mechanisms that are stable over at least a decade. In addition we have found: i) that the variations 
in the
photospheric absorption lines are confined to the
blue wing of the line (centred on velocity 
$\sim -9$ km s$^{-1}$): we tentatively ascribe this to 
an orbiting hotspot in the disc which is obscured by a disc
warp during its receding phase. ii) The wind period is manifested not only in blue-shifted 
H$\alpha$ absorption (as found by \citet{herbig03}) but also
in red-shifted emission of H$\alpha$ and H$\beta$, as well in blue-shifted absorption
of 
Na\,\textsc{i} D, Li\,\textsc{i} and Fe\,\textsc{ii} $\lambda$5018. iii) We find that the periodic
modulation of blue-shifted H$\alpha$ absorption (at a 
velocity of around
$-100$ km s$^{-1}$) is phase lagged with respect to variations in the other
lines by around $1.8$ days. This is consistent with a picture in 
which
variations at the wind base first affect chromospheric emission and then
low velocity blue-shifted absorption, followed - after a lag equal to 
the propagation time of disturbances across the wind's acceleration
region - by a response in high velocity blue-shifted absorption. Such
arguments 
constrain the size of the acceleration region to be
$\sim 10^{12}$ cm. We discuss possible mechanisms for periodic
variations within the innermost $0.1$ AU 
of the disc, including
the possibility that these variations indicate the presence of
an embedded hot Jupiter in this object.

\end{abstract}


\section{Introduction}

Although FU Orionis systems only represent a small class of young stellar objects, the benefits of understanding the physical processes 
present in these systems have implications for the whole of not only star, but also planet formation. FU Orionis systems are unique objects 
which provide the only opportunity to directly observe an accretion disc around a young star. Originally identified as a sub-set of 
pre-main-sequence stars entitled `FUors' by \citet{ambartsumian71} and \citet{ambartsumyan71a}, they are characterised by a large increase in optical 
brightness of $\sim 4$ magnitudes or more over a period of $1$--$10$ years, with a slow decline of $50$--$100$ years \citep{herbig66,herbig77}. 
Other defining properties of the class include spectral types of late F to G in the optical, with infrared spectra showing K-M 
giant-supergiant type features \citep{mould78,hartmann85,hartmann87a,hartmann87b,kenyon88,hartmann95}. FU Orionis systems are associated 
with young T-Tauri stars, evident from the strong Li\,\textsc{i} $\lambda$6707 absorption observed by \citet{herbig66}, a 
characteristic of young stars. The discovery of the second class member, V1057 Cyg, for which a pre-outburst spectrum resembled 
that of a T-Tauri star \citep{herbig58} supported this association with young stars. All FU Orionis objects are associated with star forming 
regions, as is evident from the ring shaped reflection nebulae first noted by \citet{goodrich87}, from the presence of infrared excess due to 
circumstellar dust, and by their general location in heavily extincted regions \citep{herbig77}. More recent developments have suggested these 
objects tend to have binary companions, (see table 1 in \citet{vittone05}), although the binary fraction is not necessarily higher than in other 
young stars. A companion candidate to FU Orionis itself was detected by \citet{wang04} at a separation of $0.5$ arcsecs. 
See \citet{hartmann93,kenyon95,hartmann96, hartmann98} for reviews.

FU Orionis outbursts are thought to be produced by long periods of rapid accretion \citep{larson83,lin85}, when the accretion 
rate is so great, $\mathrm{\sim 10^{\mathrm{-4}}M_{\odot{}}yr^{\mathrm{-1}}}$ \citep{kenyon88}, the inner disc heats up through viscous 
dissipation and the mid plane temperature exceeds that of the photosphere. Under these circumstances, the disc produces an absorption line 
spectrum with lower surface gravity than the central star, and the star's luminosity is swamped by factors of $100$--$1000$ \citep{hartmann96}. 
Consequently, \citet{hartmann85,hartmann87a,kenyon88,welty92,popham96,zhu09a} showed the spectrum of FU Orionis is well fitted by the 
predictions of a broad band steady state accretion disc model. The periods of rapid accretion are thought to be triggered by 
thermal instabilities in the disc associated with partial ionisation of disc material, which causes the release of dammed up material on to the 
central star \citep{hartmann85}. The trigger mechanisms are discussed by 
\citet{pringle76,hartmann85,lin85,clarke90,bell95,clarke96,lodato04,zhu09b}, and references therein. The frequency of FU Orionis events observed 
within the stellar neighbourhood has led to the belief that many low mass stars may undergo multiple FU Orionis events \citep{herbig77,hartmann85}. 
It is currently thought that up to $\mathrm{5-10}$ per cent of the matter accreted on to young T-Tauri stars occurs through repetitive periods 
of rapid accretion in FU Orionis outbursts \citep{hartmann87a,hartmann93}. This highlights the importance of understanding this episodic 
accretion process in the context of stellar formation.

The high accretion rate in FU Orionis objects is accompanied by high mass-loss rates with most objects associated with molecular outflows, optical jets 
and HH objects \citep{reipurth90,evans94}. The mass-loss rate is around $\mathrm{\sim 10^{\mathrm{-5}}M\odot{}yr^{\mathrm{-1}}}$ 
in FU Orionis itself \citep{croswell87,calvet93,hartmann95}. The strong bipolar outflow has a terminal wind velocity of $\sim 300$ km s$^{-1}$, which has been 
detected previously in the deep, broad blueshifted components of H$\alpha$ and Na\,\textsc{i} D \citep{herbig77,bastian85}. It is thought 
the outflow arises from the surface of a Keplerian disc, \citep{hartmann95}, with the possibility of a disc `chromosphere' first 
suggested by \citet{croswell87} to explain the detection of H$\alpha$ emission, and modelled in the line profiles by \citet{dangelo00}. The 
mechanism producing the rapid acceleration of material into these powerful outflows in circumstellar discs is not entirely clear. Possible models 
involve the accretion flow spinning up the star close to break up and launching magneto-centrifugal winds along opened up field lines anchored in 
the star, \citep{shu94}, or conical winds driven by the pressure gradient of the azimuthal magnetic field component and launched along opened up 
stellar field lines, \citep{konigl11}. However magnetic fields anchored in the rotating disc itself could also centrifugally accelerate 
material outwards as described by \citet{blandford82}. The detection of a $\sim 1$ kG magnetic field directed towards the observer at 
$0.05$ AU in FU Orionis by \citet{donati05} supports the involvement of magnetic fields, yet their topology and role in driving mass-loss remains unclear.

In addition to the variability on long time-scales, it appears FU Ori also exhibits unexplained variability on shorter time-scales. 
Variability in H$\alpha$ emission was first noted by \citet{herbig66} over a time-scale of one year, which was confirmed in absorption and 
emission by \citet{herbig77}. This variability was observed to exist on a monthly time-scale in observations of H$\alpha$ and Na\,\textsc{i} D  by \citet{bastian85}, 
who concluded the mass outflow activity from FU Orionis objects was a long lasting active state. 
This was confirmed by \citet{croswell87} who went on to suggest the maximum size of the region producing the variation observed in the 
absorption profiles of H$\alpha$ at $\sim 300$ km s$^{-1}$ was roughly twice the optical photospheric radius, or 
$\sim 3\times\mathrm{10}^{\mathrm{12}}$ cm. \citet{hartmann85} noted variability in the blue spectral region and in the Mg\,\textsc{i} 
$\lambda$5183 line and attributed the cause to variation in the mass ejection, concluding the wind arises from the optical disc surface 
in FU Orionis. Random, small-amplitude photometric variations on time-scales of $1$ day were observed by \citet{kenyon00}, and the rapid variations 
were attributed with the flickering of the inner accretion disc in FU Orionis.

The presence of periodicity was first evident in the brightness fluctuations after the outburst, where Fourier analysis 
by \citet{chochol80} detected periods of $1645$ and $3760$ days from observations between 1936--1976. Smaller cyclic variations were then detected 
from photometry taken in 1984--1985 by \citet{kolotilov85}, which showed a brightness periodicity in the {\it V}-band of $18.35$ days for two months in 1984 and 
interestingly, a periodicity of approximately half, $9.19$ days for the remainder of the observations. The shorter period was confirmed by 
\citet{ibragimov93}, who in addition detected a period of $9.2$ days in the {\it V}-band lightcurve from 1987 August -- 1988 January. 
\citet{kenyon88} highlighted variability in FU Orionis below 4000\AA, which maybe attributed with the periodicity found by \citet{kolotilov85}, 
or could be due to fluctuations in airmass. Further confirmations of periodicities were found in the high resolution spectroscopy 
($13$ km s$^{\mathrm{-1}}$) of $20$ nights observed from 1997--1999 analysed by \citet{herbig03}, who not only found variation in H$\alpha$ on 
daily time-scales, but also confirmed a period of $13$ to $18$ days, which peaked at $14.847$ days in the equivalent width of H$\alpha$ 
between $-110$ and $-270$ km s$^{\mathrm{-1}}$. \citet{errico03} attributed this period to a rotationally modulated wind structure, envisaging a 
non-axisymmetric magnetic structure co-rotating with the star. Note that whereas \citet{herbig03} found a periodic signal only in the 
absorption component of H$\alpha$, \citet{vittone05} detected periodicity in the H$\alpha$ emission component of $6.70$ days in 1998--2000, 
but found no correlation and no periodicity present in the absorption component of H$\alpha$. 

It is not only the strong blueshifted absorption lines affected by mass-loss that show periodicity in FU Orionis. \citet{herbig03} analysed 
spectral regions of weak photospheric lines, without strong blueshifted components (5540--5640 and 6320--6440 \AA) through cross 
correlation with a G0 Ib reference star ($\beta$ Aqr) over $20$ nights between 1997 to 1999. This analysis demonstrated that the radial 
velocity of the peak of the cross correlation function undergoes periodic variations with a period of $3.542$ days. \footnote{Variability 
also appears to be present in the 6170 \AA{} spectral region of FU Orionis, which contains weak absorption lines from the photosphere of the 
disc, however no time series analysis was conducted \citep{hartmann87a}.} It was suggested that this periodicity could be attributed to a 
non-axisymmetric hotspot, \citep{herbig03,errico03}. In the absence of any mechanism to sustain this hotspot against shear, a hotspot of radial 
extent $\delta R$ would be sheared out over $R/\delta R$ orbits. Thus it is of interest to discover whether this periodicity is sustained over many
years. Long-lived periodic behaviour would be consistent, for example, with non-axisymmetric disc structure induced by a stable magnetic
configuration or even an embedded planet \citep{clarke03}.

In this paper we present observations taken over $21$ nights from the {\it SOPHIE} \'{e}chelle spectrograph spanning a time frame of $38$ nights in 
2007. The dense time coverage and the high resolution of $4$ km s$^{\mathrm{-1}}$ has allowed the periodicities in FU Orionis to be studied 
in great detail. We observe the stability of the periodicities due to non-axisymmetric structure detected by \cite{herbig03} over a 
decade and highlight the necessity of a mechanism which maintains such a structure against shear. In section \ref{sec:obs} the details 
of the observations and data reduction techniques are described, with the variability in the line profiles described in sections 
\ref{sec:balmerlines}, \ref{sec:NaD} and \ref{sec:metallic}. The periodicity in the cross correlations of spectral regions containing weak photospheric 
lines is examined in section \ref{sec:ccf}. Possible mechanisms to sustain the periodicities over decades are discussed 
in section \ref{sec:dicus}.

\section{Observations and Data Reduction}
\label{sec:obs}

Optical spectra of FU Orionis were obtained from the {\it SOPHIE} (Spectrograph pout l'Observation des P\'{e}nom\`{e}nes des Int\'{e}rieurs 
stellaires et des Exoplan\`{e}tes) cross-dispersed \'{e}chelle spectrograph mounted on the cassigrain focus of the $1.93$-m telescope at
the Observatoire de Haute-Provence (OHP, France), covering the wavelength range 3872-6943 \AA{}. The observations were obtained using the 
high resolution mode (R$=75000$), over $21$ nights, providing an excellent opportunity to study the optical line variability 
of this object in detail. A reference spectrum of $\beta$ Aqr was also taken from {\it SOPHIE} in the same mode and used as the template in the 
cross-correlations. This G0Ib supergiant is used as the standard as it has been shown by \citet{hartmann85} that G spectral types provide 
a better signal-to-noise correlation than K spectral type templates, and it has the same effective temperature as the region of the disc 
of FU Orionis visible at optical wavelengths, as described by \citet{croswell87}. The data were reduced automatically by the pipeline, which 
includes bias subtraction, optimal order extraction, cosmic ray removal, corrections for flat fielding and wavelength calibration. The 
$39$ \'{e}chelle orders were divided by the blaze function, reconnected and barycentrically corrected, resulting in a re-sampled spectrum 
with a constant wavelength step of $0.5$ km s$^{\mathrm{-1}}$, smaller than the original spectral sampling. Details of the observations are 
shown in Table \ref{tab:observations}.

\begin{table}
  \caption{Observations of FU Orionis and $\beta$ Aqr.}
\label{tab:observations}
  \begin{tabular}{lcccc} \hline\hline
 Object & Date & Sequence & Exposure & Barycentric JD \\
&&ID&Time (s)&-2,454,000.\\
  \hline
FU Orionis & 03-01-2007 & 220020 & 1800.0 & 104.31945010\\
&04-01-2007 & 229700 & 2000.0 & 105.33395783\\
&05-01-2007 & 235720 & 2500.0 & 106.30874292\\
&06-01-2007 & 244940 & 2200.0 & 107.30133838\\
&           & 255950 &  777.0 & 107.32028452\\
&08-01-2007 & 258300 & 2000.0 & 109.31690541\\
&12-01-2007 & 274990 & 2000.0 & 113.31687568\\
&13-01-2007 & 275710 & 2000.0 & 114.31604553\\
&14-01-2007 & 276540 & 1800.0 & 115.28749519\\
&           & 276550 & 1800.0 & 115.31044553\\
&15-01-2007 & 277830 & 2000.0 & 116.37177376\\
&17-01-2007 & 279210 & 2000.0 & 118.31951493\\
&18-01-2007 & 281980 & 2000.0 & 119.42023451\\
&19-01-2007 & 282860 & 2000.0 & 120.31694694\\
&20-01-2007 & 284260 & 2000.0 & 121.33335079\\
&26-01-2007 & 294910 & 2000.0 & 127.37686294\\
&27-01-2007 & 295440 & 2000.0 & 128.30713983\\
&28-01-2007 & 296290 & 2000.0 & 129.30472606\\
&29-01-2007 & 297200 & 2000.0 & 130.29181404\\
&30-01-2007 & 298190 & 2000.0 & 131.30359823\\
&31-01-2007 & 299070 & 2000.0 & 132.30679379\\
&05-02-2007 & 303430 & 1800.0 & 137.33476702\\
&10-02-2007 & 306160 & 1800.0 & 142.33284243\\
$\beta$ Aqr& 08-10-2007 & 465830 & 300.0  & 382.34761084\\
  \hline
  \end{tabular}
\end{table}

\begin{figure*}
 \centering
\includegraphics[angle=0,width=0.99\linewidth]{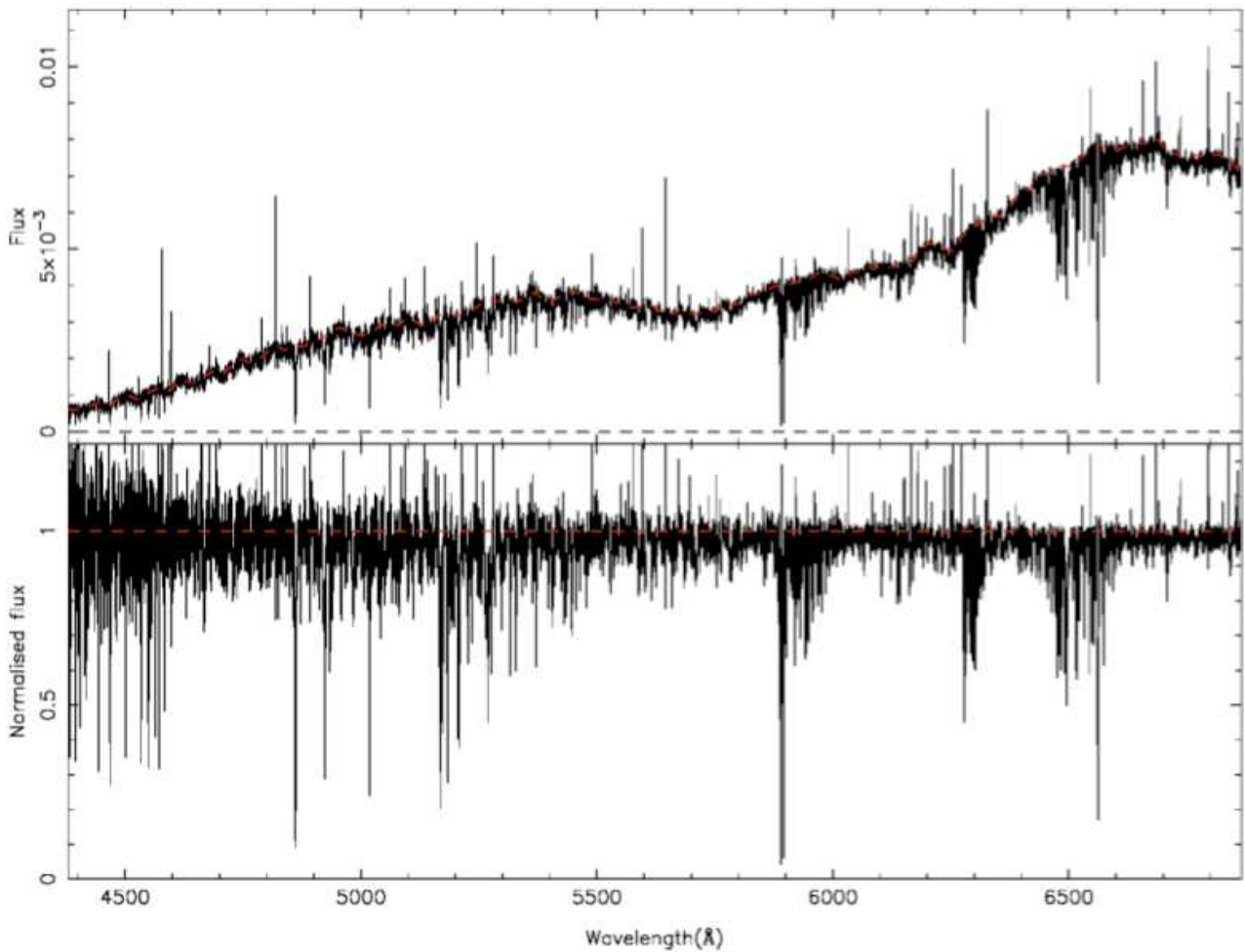}
\caption{High resolution {\it SOPHIE} \'{e}chelle spectra mounted on the $1.93$-m telescope at the OHP taken on JD=2,454,106.309. The top panel shows the 
original spectrum, with the red dashed line representing the continuum level. The bottom panel shows the spectrum normalised to unit continuum.}\label{fig:normalise}
\end{figure*} 

The data were continuum subtracted by clipping out the spectral features and performing median and boxcar filtering over a scale length of 
$20$\AA{}, to obtain a pure continuum spectrum, which was then subtracted from the featured spectrum and normalized to unit continuum. See 
\citet{battaglia08} for more details on this method. This was performed over the wavelength range 4100--6865 \AA{} to ensure variations in 
the continuum were not included in the resultant line profiles and to minimize the propagation of errors dues to edge effects. Fig. 
\ref{fig:normalise} shows the initial normalisation of the JD=2,454,106.309 observation over this region. Variance in the continuum 
was minimized by averaging all $21$ nights of the normalised spectra and subtracting this from each spectra individually to leave the residual 
spectra. These residuals were then continuum subtracted again and the result was re-combined with the initial average spectrum to ensure 
fluctuations in the continuum do not affect the line profile variability.

Spurious emission was removed by flagging the points which were above three sigma from the mean, both in each spectra individually, 
and across simultaneous points in all $21$ observations, and replacing with the result of a median filter across a region of $\pm0.1$ \AA{} 
around the bad value. This ensured large variations in emission over the spectral series were removed, reducing the potential contribution 
from spurious lines. Telluric lines were identified through visual comparison between all the unbarycentrically corrected spectra stacked 
with the median of all spectra. Aligned, narrow lines which persisted in the median spectrum were clipped and removed from further analysis.

The continuum subtracted spectra were cross-correlated with a template G0Ib reference star, namely $\beta$ Aqr taken from the same 
instrument, via the methods described in \citet{tonry79}. The template was shifted to zero velocity, as it has an radial velocity 
measured to be $7.11$ km s$^{-1}$, when cross correlated with a synthetic spectrum of a G0 star with T$_{\mathrm{eff}}=6000$ K, log(g)$=3.0$. 
This specific model was chosen as it provided the best match to the observed line strengths in the $\beta$ Aqr spectrum. The radial 
velocity was derived by fitting a Gaussian to the peak of the symmetric cross-correlation function. This was necessary as previous 
measurements were made with far less accurate resolution by \citet{wilson53}. The observed template was continuum subtracted in the same manner 
as described above. Furthermore, the synthetic stellar spectrum was smoothed using a Gaussian with full width half maximum (FWHM), comparable to 
the corresponding unblended, unsaturated absorption lines in the relevant regions of the $\beta$Aqr spectrum. The resulting spectrum was used as
an addition reference standard in cross correlations in order to confirm the zero point of the spectral lines in the $\beta$ Aqr spectrum.
Using this method over a spectral interval not only improves the signal-to-noise, but also allows for the effects of line blending 
or overlap and reduces scatter in the results \citep{zhu09a}, ideal for the weaker metallic lines.

Errors were obtained for the cross correlations through estimating the variance and hence signal-to-noise in each spectral region, using the areas 
identified though spectral clipping to be pure continuum. The covariance matrix across $3$ adjacent pixels in these regions were 
calculated and the noise reduction factor was computed. This was combined with the original estimate for the signal-to-noise, which 
accounted for the noise correlated across nearby pixels in a re-binned \'{e}chelle spectrograph. Random amounts of Gaussian noise 
centred on the signal-to-noise value were then added to the spectra, which were then cross-correlated with the template spectrum once more. 
The resulting variance in the cross correlations and line profiles was calculated and the $3\sigma$ errors found for each pixel, allowing
an error bar to be associated with each point.

\section{Balmer Lines}
\label{sec:balmerlines}

\subsection{H$\alpha$ Line Profiles}
\label{sec:halphalines}
\begin{figure}
\centering
\includegraphics[width=0.83\linewidth]{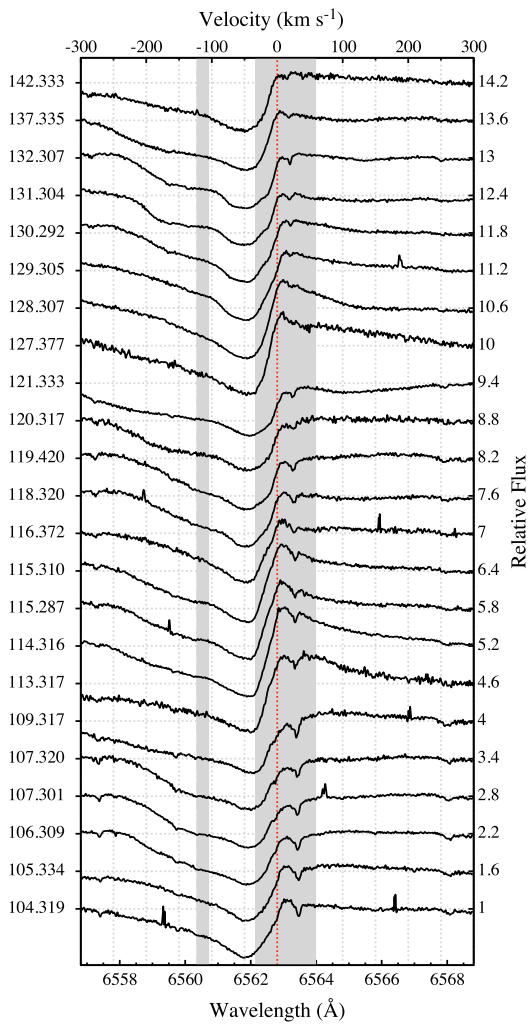}
\caption{Variations in the H$\alpha$ profile of FU Orionis obtained from the {\it SOPHIE} \'echelle spectrograph mounted on the $1.93$-m telescope at the OHP.
The $21$ spectra cover a time frame of $38$ nights, with the Julian dates of the spectra -2,454,000 displayed on the y-axis. All spectra have a resolution
of $4$ km s$^{\mathrm{-1}}$, are barycentrically corrected, and placed in the rest frame of FU Orionis. The shaded region depicts the velocities where periodicity 
was found to be significant below the $0.01$ FAP level.}
\label{fig:halphaprof}
\end{figure}

Fig. \ref{fig:halphaprof} shows the H$\alpha$ line profiles obtained from {\it SOPHIE} from 2007 January 3$^{\mathrm{rd}}$ -- February 10$^{\mathrm{th}}$. 
All profiles appear P Cygni in nature. The profiles show a blueshifted absorption component with the edges extending up to around 
$-250$ to $-300$ km s$^{-1}$ with the
deepest absorption at $\sim -50$ km s$^{-1}$ and strength weakening as blueshifted velocity increases. The redshifted emission component shows 
greater variability in strength: only some profiles show a redshifted emission component as first noted by \citet{croswell87}, with the 
red edge of the emission extending from $200$ to $250$ km s$^{-1}$, similar to those observed by \citet{bastian85}. The presence of a correlation 
between the variability of the strength of the redshifted emission peak and the width of the blueshifted absorption trough is initially 
not entirely clear, although \citet{dangelo00} suggested a negative correlation. \footnote{A variation in the minimum of the blueshifted absorption 
may also give rise to this variability. However, the superposition of H$\alpha$ emission components make this difficult to identify.} Fig. 
\ref{fig:halphaprof} shows no evidence a negative correlation is sustained in these data, for example on JD=2,454,113.317 there is strong redshifted 
H$\alpha$ emission, combined with broad blueshifted absorption.

\begin{figure}
 \centering
 \includegraphics[width=0.99\linewidth]{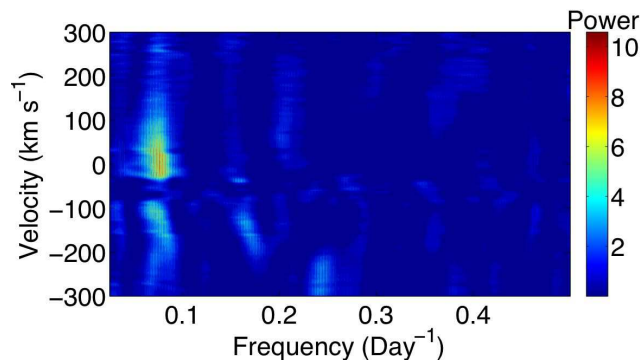}
 \caption{Contour periodogram showing the significant periods found in the H$\alpha$ line profiles. The $0.01$ FAP corresponds to power \textgreater $7.68$.}
 \label{fig:halphaperiod}
\end{figure}

To identify any quantifiable relationship between the variation of the absorption and emission components of the observed P Cygni profiles, 
a light curve was constructed for each pixel (every $0.5$ km s$^{-1}$) in the line profile, which 
were then tested for 5000 different periods between $2$--$40$ days, evenly distributed in frequency space, using the method initially described by 
\citet{scargle82}, and revised by \citet{horne86}. The Lomb-Scargle periodogram is based on least-squares sine curve fitting. Consequently, 
it is an ideal tool for these data because in addition to robustness to irregular coverage it also enables longer periods to be explored and is 
commonly used to explore up to the approximate total coverage period of the data. The significance of the periods were found to false alarm probability, 
(FAP) of less than $0.01$ using Fisher's method of randomization \citep{linnell85}. The periodogram showing the significant periods found in the H$\alpha$ 
line profile are shown in Fig. \ref{fig:halphaperiod}. The periodogram highlights the significant periods with FAP\textless $0.01$, corresponding to a power 
\textgreater{} $7.68$. These were in two distinct regions from $-122.0$ to $-105.5$ km s$^{-1}$ and from $-32.5$ to $+85.5$ km s$^{-1}$, shown by the shaded 
regions in Fig. \ref{fig:halphaprof}. The range of significant periods detected was from $11.59$--$16.29$ days,
with the most probable value found from the `centre of power' of all the significant powers detected in the line profile at $13.28$ days.
This period is similar to that found by \citet{herbig03}, who detected a period in the `fast wind' in the equivalent width, (EW) of the P Cygni 
absorption between $-110$ to $-270$ km s$^{-1}$ of $14.847$ days. However, unlike \citet{herbig03}, we detect the period in the line profile 
itself, and not in the EW of the absorption component. The overlap in velocity space of the significant periods detected with the results of 
\citet{herbig03} confirms the stability of the period in the blueshifted absorption component of H$\alpha$ over a decade. In addition, Fig. 
\ref{fig:halphaperiod} extends the detection of the significant period to less negative velocities than the results of \citet{herbig03}. The absence 
of periodicity from $-105.5$ to $-32.5$ km s$^{-1}$, suggests the periodicity detected \textless $-100$ km s$^{-1}$ is due to variations in the FWHM of 
the blueshifted absorption component of H$\alpha$ as proposed by \citet{dangelo00}. This `absent' region in Fig. \ref{fig:halphaprof} 
corresponds to the deepest blueshifted absorption on all nights observed. Likewise, Fig. \ref{fig:halphaprof} also demonstrates less variability 
around the minimum of the absorption trough, which could be attributed to the suggestion by \citet{croswell87} that the Balmer lines are so 
optically thick that they are saturated, and consequently are not very vulnerable to variations in physical parameters such as density/temperature 
fluctuations or changes in the mass-loss rates. However, at higher negative velocities, further out in the expanding wind, the H$\alpha$ line may no longer 
be saturated, and environmental variations are reflected in the variability of the line profiles in such regions. Alternatively, the phase-complexity 
of the superposition of different variable components contributing to the H$\alpha$ profile may prevent any periodicity in this region from being 
detected. The periodicity in the strength of the redshifted emission component is the same as the periodicity in the blueshifted 
absorption component, shown by the alignment in frequency of the significant peaks in the periodogram in Fig. \ref{fig:halphaperiod}. This is 
the first time such correlation has been detected. Previous observations by \citet{vittone05} have seen periodicity in the redshifted emission 
component of $6.70$ days, which is not evident in our results. The dense time coverage of our results confirms this period is no longer present, 
and could be a result of aliasing in the earlier observations. Therefore, in contrast to previous beliefs, the periodic variability in FU Orionis seems to 
be stable over hundreds of periods, because the period detected here agrees well with the significant periods of $13$--$18$ days found by \citet{herbig03}.

\begin{figure*}
 \centering
 \includegraphics[width=0.95\linewidth]{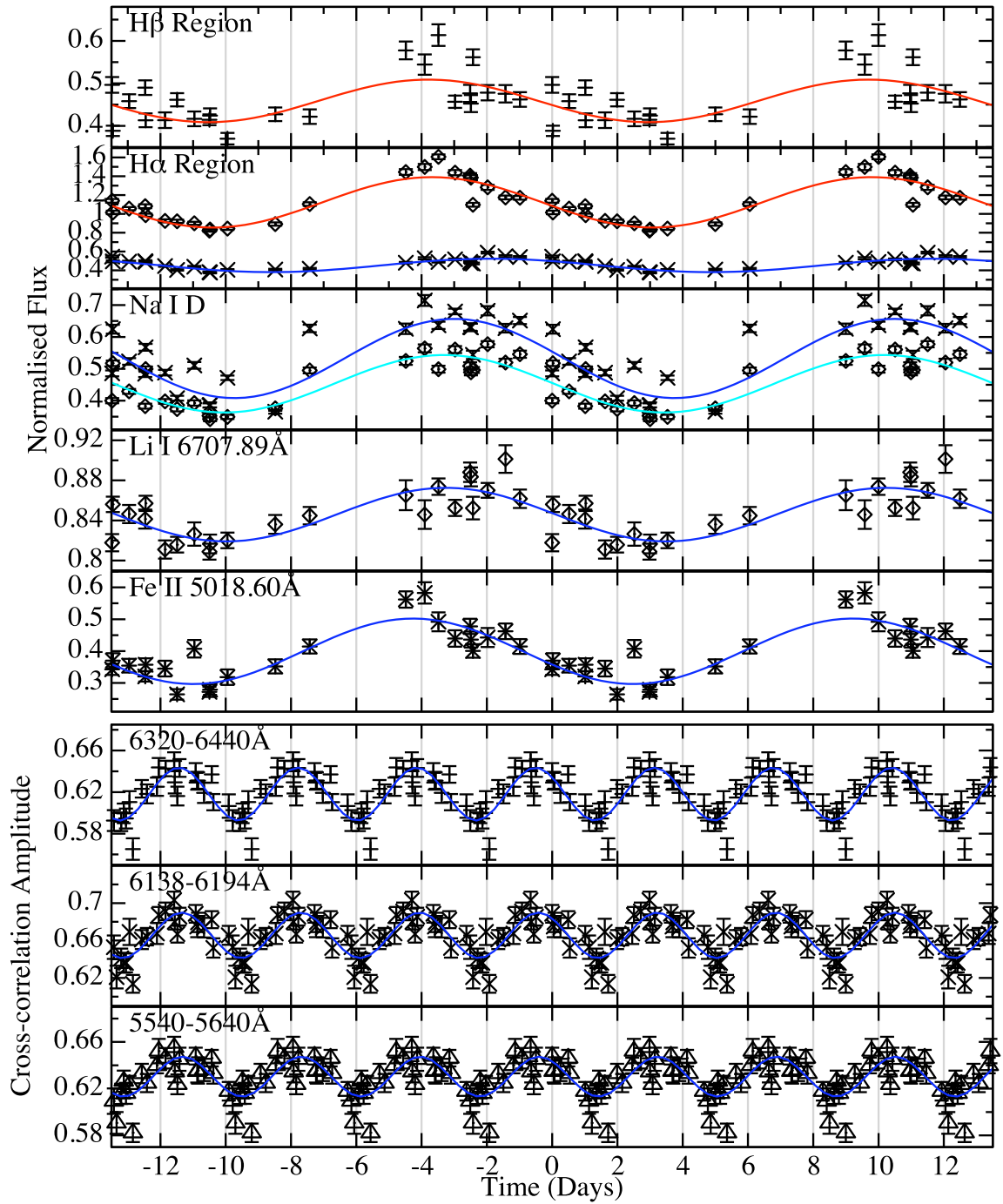}
 \caption{Phase diagram comparing the periodicity found in different spectral lines (top five panels), with the periodicity found in the cross 
correlations, (bottom three panels). The spectral lines are phase folded with the variance weighted total average period of all five regions, which is 
$13.48$ days. The spectral lines are plotted at the power weighted average velocity of the relevant significant detections (from top to bottom), H$\beta$ at 
$+3.5$ km s$^{-1}$, H$\alpha$, (in redshifted emission at $+10.0$ km s$^{-1}$ plotted with diamonds, and in blueshifted absorption at $-109.5$ km s$^{-1}$ 
plotted with crosses), Na\,\textsc{i} D, (with Na D$_{1}$ $\lambda$5895.92 at $-10.0$ km s$^{-1}$ plotted with crosses and Na D$_{2}$ $\lambda$5889.95 at $-11.0$ 
km s$^{-1}$, plotted with diamonds), Li\,\textsc{i} $\lambda$6707 at $-35.0$ km s$^{-1}$ and Fe\,\textsc{ii} $\lambda$5018 at $-39.0$ km s$^{-1}$. 
The regions cross correlated are shown in the bottom three panels. The cross correlation profiles were phase folded with a variance weighted combination 
of the individual periods, which averaged $3.64$ days. The cross correlations are also plotted at the power weighted average velocity which are, (from
top to bottom) $-9.0$ km s$^{-1}$ in 6320--6440\AA{}, $-7.5$ km s$^{-1}$ in 6138-6191\AA{} (from cross correlations with observed template of 
$\beta$Aqr) and $-12.0$ km s$^{-1}$ in 5540--5640\AA{} (from cross correlations with synthetic stellar spectrum of G0 star at T$_{\mathrm{eff}}$=6000 K, 
log(g)$=3.0$). 
The solid lines show the sinusoidal line of best-fitting for each individual detection, which are plotted in blue when the period is detected at 
blueshifted velocities (the fit to Na D$_{\mathrm{2}}$ is plotted in light blue for clarity), and red corresponding to detections in redshifted velocities. 
All error bars are shown to the $3$ sigma level.}
 \label{fig:phase}
\end{figure*}
 
The phase diagram in Fig. \ref{fig:phase} shows the phase relationship between the periodic behaviour detected in various spectral regions phase folded 
with the variance weighted average period of $13.48$ days from all line profiles investigated. In each 
case the light curve is constructed at the power weighted centre of velocity for the periodic region concerned. The 2$^{\mathrm{nd}}$ panel from the top 
shows the two periodic regions of the H$\alpha$ profile. The amplitude of variation corresponding 
to the oscillations in redshifted emission in the  H$\alpha$ profiles (the red line) is greater than those at the faster wind velocity (the blue line). 
This is expected from the appearance/disappearance of the H$\alpha$ emission peak, in contrast with the broadening-narrowing FWHM of 
the blueshifted absorption in Fig. \ref{fig:halphaprof}. Fig. \ref{fig:phase} confirms the `absent' region in Fig. \ref{fig:halphaprof} is not 
simply lacking of periodicity due to complex superposition of the different variable components, since this would require the relative 
contributions to be equal and opposite, which is clearly not the case. The saturation of the H$\alpha$ line between $-105.5$ to $-32.5$ km s$^{-1}$ seems a more
plausible explanation. The sinusoidal curve fitted to the fast blueshifted data also appears to have a slight lag of $\sim 1.80$ days in comparison with 
the sinusoidal curves fitted to the data corresponding to oscillations in the redshifted emission peak. This could stem from the different formation 
locations of the two line components, with the redshifted emission formed in the disc chromosphere, and the blueshifted absorption formed further out in the 
expanding wind.

\subsection{H$\beta$ Line Profiles}
\label{sec:hbetalines}

\begin{figure}
\centering
\includegraphics[width=0.83\linewidth]{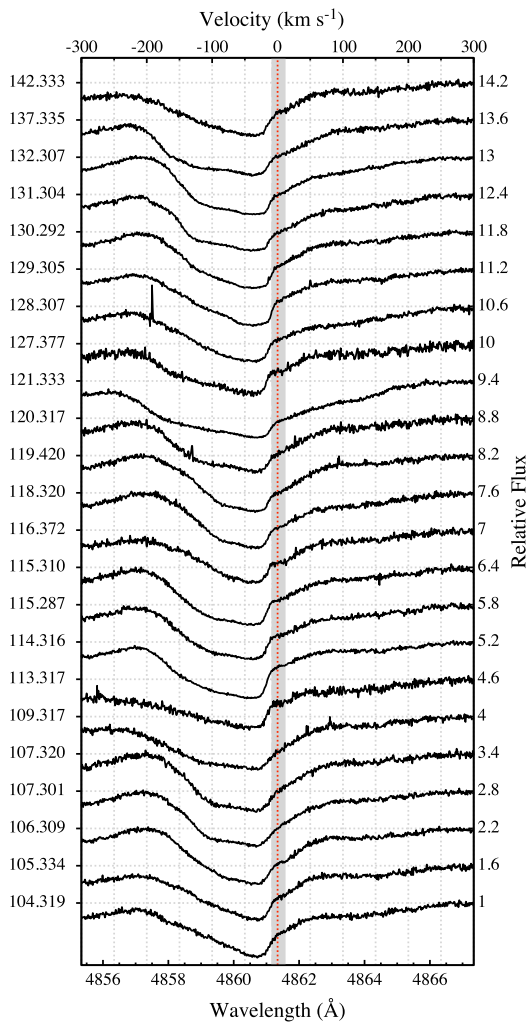}
\caption{Same as Fig. \ref{fig:halphaprof} except for the H$\beta$ profile at 4861.34 \AA{}.}
\label{fig:hbetaprof}
\end{figure}

\begin{figure}
 \centering
 \includegraphics[width=0.99\linewidth]{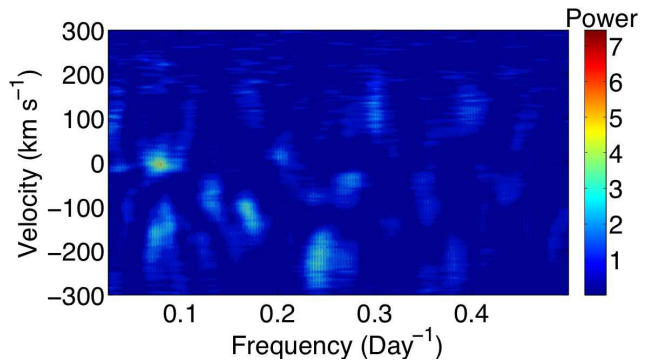}
 \caption{Contour periodogram showing the significant periods found in the H$\beta$ line profiles. The $0.01$ FAP corresponds to power \textgreater{} $7.02$.}
 \label{fig:hbetaperiod}
\end{figure}

The H$\beta$ line profiles obtained from {\it SOPHIE} are shown in Fig. \ref{fig:hbetaprof}. All profiles show broad asymmetric absorption troughs, with 
stronger absorption on the blueshifted edge, which extends out to faster velocities than the redward absorption wing. This is to be expected for the 
H$\beta$ profile, which is thought to be produced in an accelerating, massive outflow \citep{hartmann94}. However, all the profiles appear more skewed 
towards blueshifted velocities than previously observed in 1992 by \citet{hartmann95}, which may imply an increase in the mass-loss rate since this 
epoch. The maximum depth of the absorption trough appears to remain consistent at $\sim -25$ km s$^{-1}$ throughout our observations. However, there is 
still significant night-to-night variability visible in Fig. \ref{fig:hbetaprof}. For instance, at slower negative 
velocities, and low positive velocities the gradient of the profile appears highly variable. This could be an effect of the varying relative absorption 
strengths between the redshifted and blueshifted components contributing to the overall line profile. In addition, the FWHM of the blueshifted 
absorption component appears to vary significantly throughout the observations, which seems to correlate with the apparent broadening of the H$\alpha$ 
absorption discussed in section \ref{sec:halphalines}. This is evident when comparing observations on JD=2,454,113.317 and JD=2,454,132.307, 
displayed in Figs. \ref{fig:halphaprof} and \ref{fig:hbetaprof}, both of which show relatively broad and narrower blueshifted 
absorption components respectively.

The presence of any periodic variability in the H$\beta$ line profiles was investigated between $2$--$40$ days with the analysis described in section 
\ref{sec:halphalines}. The region of frequency-velocity space identified as significant to below the $0.01$ FAP, which represents a power \textgreater{} 
$7.02$ in these data, corresponds to a period from $8.82$ to $14.68$ days at velocities of $-8.0$ to $+11.0$ km s$^{-1}$ in the line profile. This feature can be 
seen as the red blob in Fig. \ref{fig:hbetaperiod}, with a power-weighted average period of $12.89$ days and at a average velocity of $+3.5$ km s$^{-1}$. This is 
the first detection of periodicity in H$\beta$. It is surprising that the periodicity is not seen at faster blueshifted velocities like H$\alpha$, given 
the formation of the both lines within the massive outflow. However, the examination of the top two panels in Fig. \ref{fig:phase}, shows that the 
periodicity in the redshifted emission of H$\alpha$ and redshifted side of H$\beta$ appear to be very similar in-phase. This could suggest that it is the 
redshifted emission component in H$\beta$, formed in the disc chromosphere, and not a variation in strength of the superimposed H$\beta$ redshifted 
absorption, which is the source of this periodicity. On examination of Fig. \ref{fig:hbetaprof} at the velocities where a period was identified 
(the grey shaded region between $-8.0$ to $+11.0$ km s$^{-1}$), it appears this cyclic 
variability corresponds to the shallowing/deepening of the `ridge' connecting the blueshifted absorption to the redshifted absorption. 
Consequently, a variation in H$\beta$ emission could produce this effect, and a narrower formation region of H$\beta$ emission in comparison to 
H$\alpha$ emission within the chromosphere, or the additional superimposed variation in the redshifted absorption component of H$\beta$ may prevent 
this periodicity being detected at faster redshifted velocities alike H$\alpha$.

\subsection{H$\gamma$ Line Profiles}
\label{sec:hgammalines}
The presence of periodic variability was also tested analogously in the H$\gamma$ profiles. In this case, the observations showed no evidence for 
periodicity in the profiles from $-300$ to $+300$ km s$^{-1}$ above the random noise level, which is higher in this shorter wavelength region.

\section{Na\,\textsc{i} D Profiles}
\label{sec:NaD}

\begin{figure}
\centering
\includegraphics[width=0.83\linewidth]{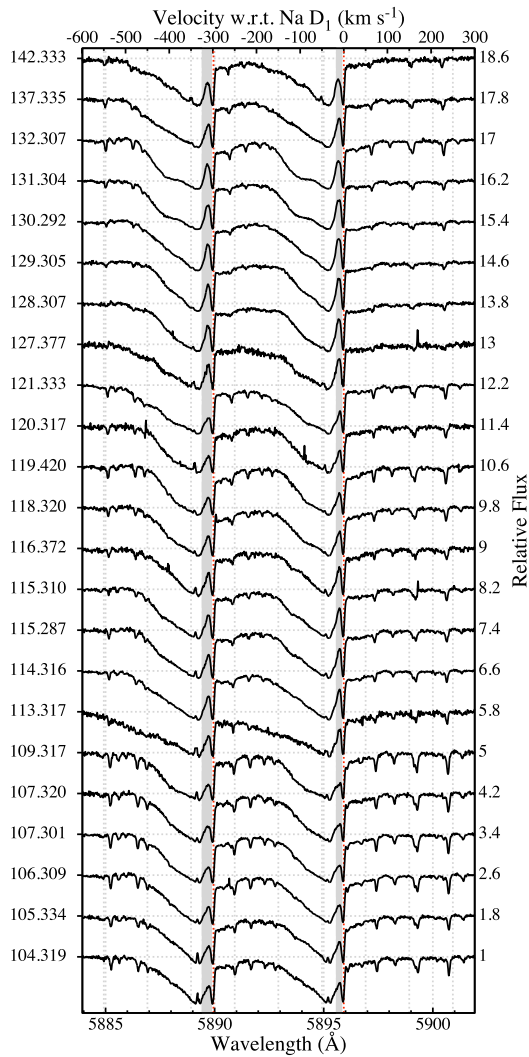}
\caption{Same as Fig. \ref{fig:halphaprof} except for the Na\,\textsc{i} D profile of FU Orionis. The lower axis is centred around Na\,\textsc{i} D$_{1}$ 
at 5895.92 \AA{}, and the upper axis is plotted with respect to Na\,\textsc{i} D$_{2}$ at 5889.95 \AA{}.}
\label{fig:NaDprof}
\end{figure}

\begin{figure}
 \centering
 \includegraphics[width=0.99\linewidth]{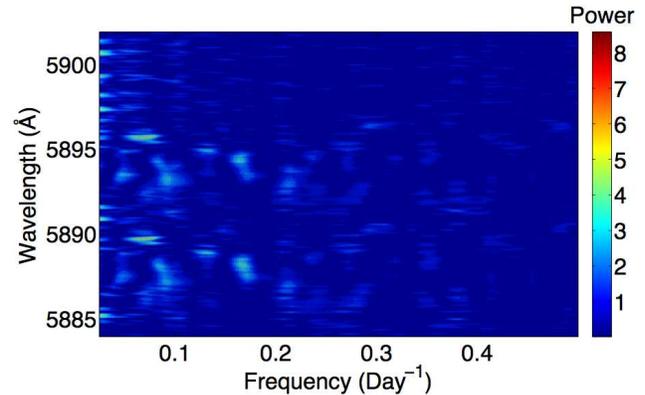}
 \caption{Contour periodogram showing the significant periods found in the Na\,\textsc{i} D line profiles. The $0.01$ FAP corresponds to power 
\textgreater{} $6.67$.}
 \label{fig:NaDperiod}
\end{figure}

The Na\,\textsc{i} profiles obtained from {\it SOPHIE} are shown in Fig. \ref{fig:NaDprof}. All profiles show complex absorption components, with no noticeable 
emission components, as first observed in 1981 by \citet{bastian85}. The blueshifted profiles both seem to contain two dominate absorption components, a 
high-velocity broad feature, superimposed with a narrow, low-velocity absorption dip \citep{croswell87}. The broad blueshifted feature appears to be variable 
during our observations with the edge extending out to velocities between $-150$ to $-250$ km s$^{-1}$, (slightly slower than the expansion velocities seen in 
H$\alpha$ in Fig. \ref{fig:halphaprof}). This is consistent with the lines' formation in a differentially expanding wind, similar to the broad blueshifted 
absorption of H$\alpha$, but in a more compact region, with the wind continuing to expand throughout the more extended H$\alpha$ formation region 
\citep{bastian85}. \citet{bastian85} went on to conclude that the temperature of the accelerating region must increase outwards, because H$\alpha$ generally 
needs higher temperatures to be formed than Na\,\textsc{i} D, yet \citet{croswell87} showed the H$\alpha$ absorption could be formed at lower temperatures 
with an increased mass-loss rate. In addition, the relatively broad absorption feature of Na\,\textsc{i} D seems to be comprised 
of two separate variable absorption components, which are visible separately on some nights and appear blended together on others, (see JD=2,454,104.319 
and JD=2,454,130.292 on Fig. \ref{fig:NaDprof}). This is consistent with \citet{croswell87} and \citet{hartmann95}, who observed a single-component and 
\citet{bastian85} who observed two-components in the broad absorption features. On the other hand, the narrow low-velocity component $\sim -3$ km s$^{-1}$ 
does not appear to vary significantly during our observations, which is consistent with it's proposed formation in an expanding circumstellar shell, at large 
distances from the object, as discussed by \citet{bastian85}.

The presence of periodicity in the Na\,\textsc{i} D line profiles between $2$--$40$ days were analysed using the same methods described in section 
\ref{sec:halphalines}. The significant periods found in this doublet are shown by the two red regions in Fig. \ref{fig:NaDperiod}, 
which are below the $0.01$ FAP, with power \textgreater{} $6.67$. In the Na D$_{1}$ line at 5895.92 \AA{}, periodicity is significant to less than the $0.01$ FAP 
between $-16.0$ to $-5.5$ km s$^{-1}$ for periods of $12.41$ to $17.87$ days. The range of significant periods detected in the Na D$_{1}$ line not only overlaps 
with the periods detected to less than the $0.01$ FAP in the Na D$_{2}$ line at 5889.95 \AA{}, but are also present at the same relative velocities in the 
line profile, shown by the shaded regions in Fig. \ref{fig:NaDprof}. Significant periodicity was detected in Na D$_{2}$ between $-25.0$ to $-5.5$ 
km s$^{-1}$ with periodicity ranging from $11.59$ to $17.54$ days. This is the first detection of periodicity in the Na\,\textsc{i} D profile and the symmetry 
of these two individual detections justified their combination to calculate an overall power-weighted average period of $14.27$ days for the Na\,\textsc{i} 
D profile. There are also some other detections in Fig. \ref{fig:NaDperiod} that appear 
to have power values above the power corresponding to the $0.01$ FAP level, however these detections are situated on the very edge of the periodogram at small 
frequencies and do not coincide with an extended feature. Therefore these `significant detections' are assumed to be a result of edge affects in the 
frequency sampling and are not considered in further analysis.

The origin of the periodicity highlighted by the grey shaded regions on the line profiles in Fig. \ref{fig:NaDprof}, is difficult to distinguish due to the 
complexity of the blueshifted absorption components contributing to both lines. However, the periodicity of both lines in the doublet appears to be cut off 
at the edge of the narrow low-velocity blueshifted absorption component. This is consistent with the formation of this low-velocity component in a slowly 
expanding shell at a large distance from the object. Therefore, the periodicity is more likely associated with the lower velocity end of the broader, more 
variable blueshifted components, which are attributed to the expanding wind. This is supported by the periodicities overlapping with the significant 
detections in H$\alpha$ and H$\beta$. The detection of these periods at relatively low velocities in comparison to the maximum velocity of the outflow 
$\sim -150$ to $-250$ km s$^{-1}$ in this case, suggests the origin of this periodicity is lower down in the expanding wind, in 
contrast to H$\alpha$ in section \ref{sec:halphalines}, where periodicity in the blueshifted absorption component was seen centred at $-109.5$ km s$^{-1}$. 
This disparity could be due to a combination of the more extended formation region of H$\alpha$ and the relatively reduced temperature sensitivity of the 
Na\,\textsc{i} D, \citep{croswell87}, which may prevent a significant periodic signal at the edges of the formation region, in the expanding wind. Fig. 
\ref{fig:phase}, supports this since the 3$^{\mathrm{rd}}$ panel from the top shows the periodicity in both lines phase folded with a variance weighted 
average overall period of $13.48$ days. Both Na\,\textsc{i} D lines, (D$_{1}$ shown in dark blue, D$_{2}$ shown in light blue) appear to have a slight lag 
with the H$\alpha$ emission, thought to be formed in the disc chromosphere, and a slight lead over the H$\alpha$ absorption, formed further out in the 
expanding wind, shown in the panel above in Fig. \ref{fig:phase}.

\section{Metallic Line Profiles}
\label{sec:metallic}

\begin{figure}
 \centering
 \includegraphics[width=0.83\linewidth]{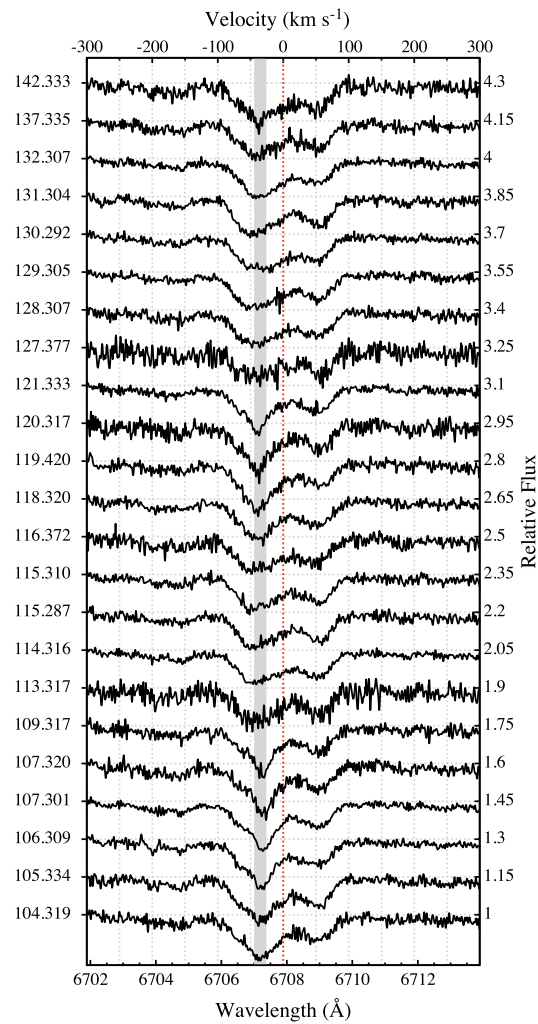}
 \caption{Same as Fig. \ref{fig:halphaprof} except for the Li\,\textsc{i} profile at 6707.89 \AA{}.}
 \label{fig:liprof}
\end{figure}

\begin{figure}
 \centering
 \includegraphics[width=0.99\linewidth]{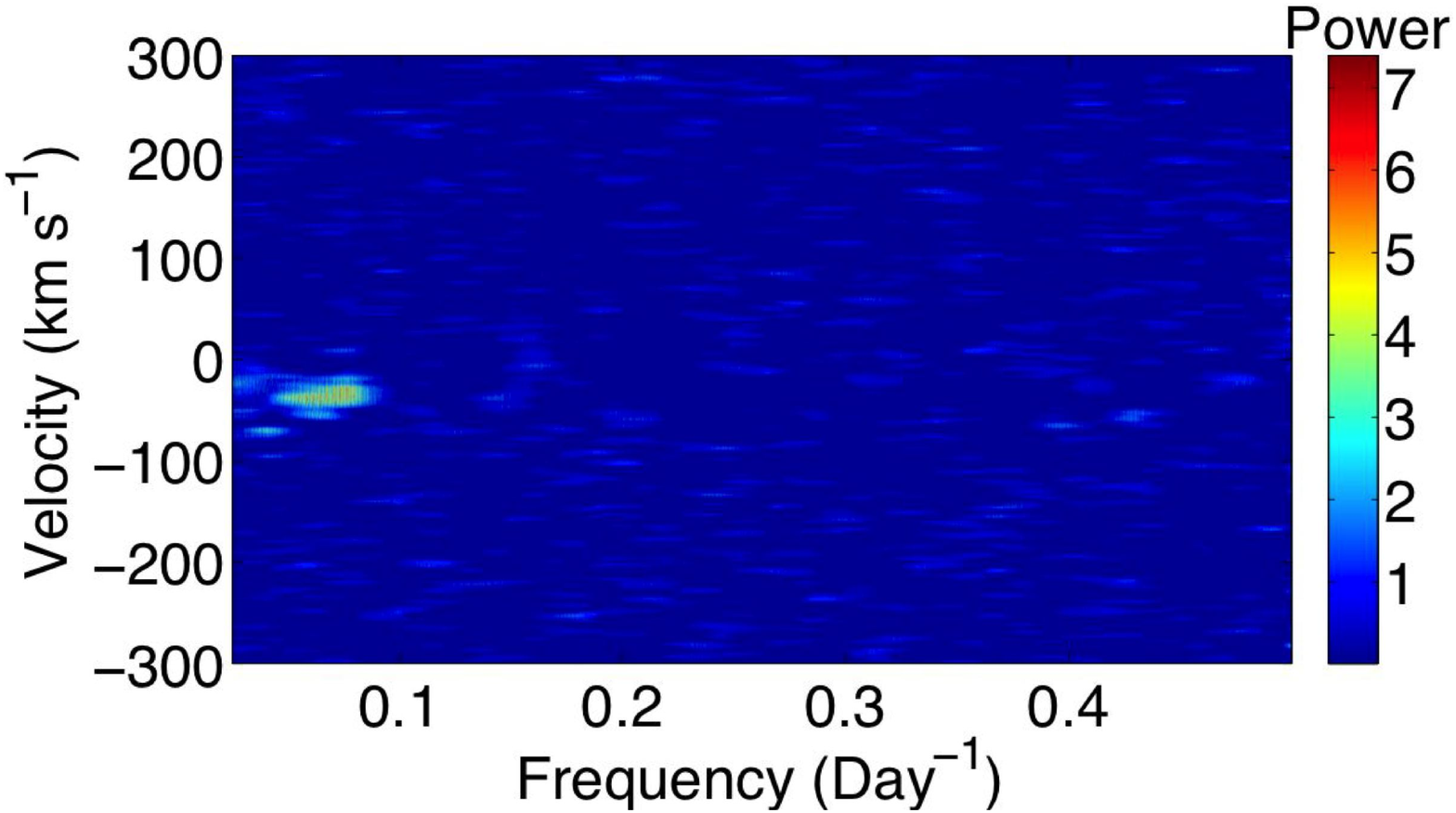}
 \caption{Contour periodogram showing the significant periods detected in the Li\,\textsc{i} line profile. The $0.01$ FAP corresponds to power 
\textgreater{} $6.76$.}
 \label{fig:licontour}
\end{figure}

\begin{figure}
 \centering
 \includegraphics[width=0.83\linewidth]{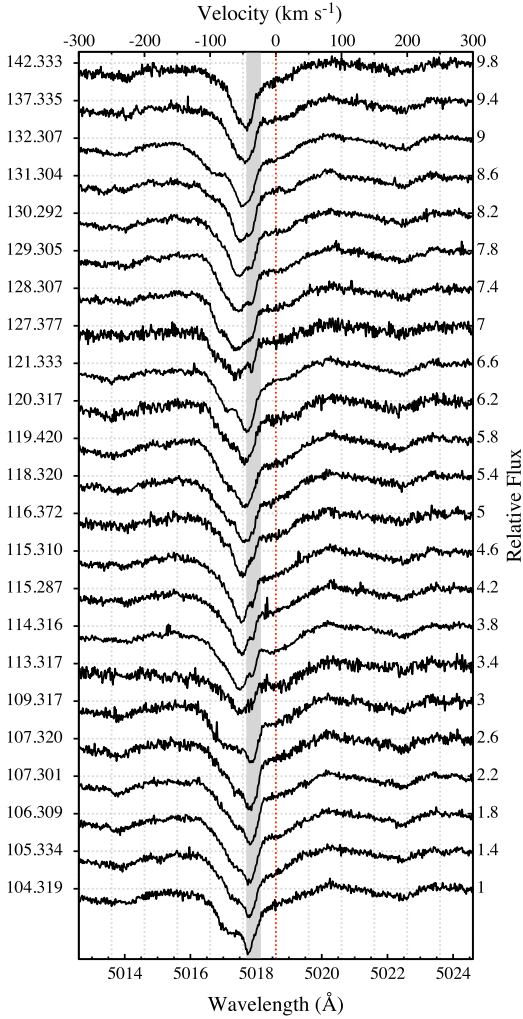}
 \caption{Same as Fig. \ref{fig:halphaprof} except for the Fe\,\textsc{ii} profile at 5018.60 \AA{}.}
 \label{fig:5018prof}
\end{figure}

\begin{figure}
 \centering
 \includegraphics[width=0.99\linewidth]{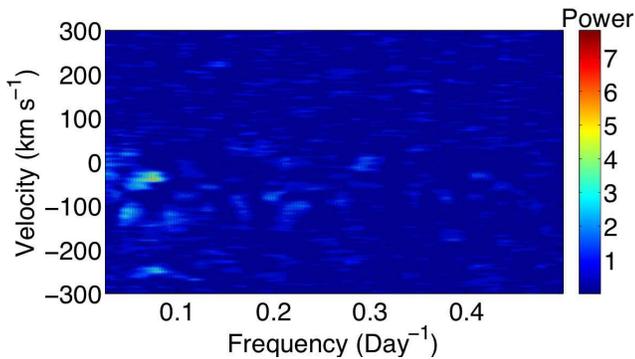}
 \caption{Contour periodogram showing the significant periods detected in the 5018 Fe\,\textsc{ii} line profile. The $0.01$ FAP corresponds to power 
\textgreater{} $6.73$.}
 \label{fig:5018contour}
\end{figure}

Line profile variations in Li\,\textsc{i}  $\lambda$6707 from 2007 January 3$^{\mathrm{rd}}$ to February 10$^{\mathrm{th}}$ are shown in Fig. 
\ref{fig:liprof}. These profiles support the accretion disc hypothesis of FU Orionis objects as described by \citet{hartmann85,hartmann87a,zhu09a}, 
since the profile displays a double peaked nature, as initially observed by \citet{herbig89}; however variability only exists in the 
blueshifted absorption. Unlike H$\alpha$, as described in section \ref{sec:halphalines}, the width of the blueshifted absorption appears to 
remain constant. The blueshifted absorption instead varies in strength, causing the double peaked line profile to appear more and less 
asymmetrical, see JD=2,454,121.333 and JD=2,454,127.377, on Fig. \ref{fig:liprof}. These observations coincide with the dates when the 
H$\alpha$ profile had no redshifted emission, and strong redshifted emission respectively (see Fig. \ref{fig:halphaprof}). This could indicate a 
link between the origin of the variability in the H$\alpha$ line and Li\,\textsc{i} $\lambda$6707.

The possibility of periodicity between $2$--$40$ days in the Li\,\textsc{i} $\lambda$6707 line profile was investigated via the method described 
in section \ref{sec:halphalines}, producing significant detections at velocities from $-43.0$ to $-27.0$ km s$^{-1}$, with periods between 
$12.21$ and $19.78$ days, resulting in a `centre of power' period at $14.81$ days. This is the first detection of periodicity in Li\,\textsc{i} and can be 
clearly identified as the large orange-red structure in Fig. \ref{fig:licontour}, which displays the resultant periodogram for the frequency-velocity 
space analysed. There appears to be an addition feature on the periodogram in Fig. \ref{fig:licontour} that stands out in yellow above the background 
at velocities between $-69.5$ to $-68.0$ km s$^{-1}$ corresponding to a period between $21.22$ and $29.39$ days, and a power-weighted average period of $24.76$ days. 
This feature is significant below the $0.01$ FAP level, which corresponds to a power \textgreater{} $6.76$. On inspection of Fig. \ref{fig:licontour}, 
it is unclear whether this periodicity is connected to the larger feature $\sim -35$ km s$^{-1}$, as it is located at slightly more negative velocities 
towards shorter frequencies. However, since there is no overlap in frequency space between the significant periods detected it is treated as a separate 
detection and for the remainder of this discussion is classed as noise fluctuations, until further detections prove the contrary.

The location of the main spike in power on Fig. \ref{fig:licontour} coincides with the deepest absorption in the Li\,\textsc{i} $\lambda$6707 
profile in Fig. \ref{fig:liprof}, located at $\sim -35$ km s$^{-1}$, with the connection highlighted by the shading of the periodic region. This confirms 
the periodicity in Li\,\textsc{i} is a result of the varying strength of the blueshifted absorption peak, and not a periodic broadening of the line profile. 
The confinement of periodicity to purely negative velocities indicates this could be another periodic wind signature, although it was previously thought 
the Li\,\textsc{i} absorption line was not formed in the wind \citep{calvet93}. Despite the apparent difference in nature of the periodicities on the blue 
wing with H$\alpha$, the periodicity observed is similar in nature to that of Na\,\textsc{i} D in Fig. \ref{fig:NaDprof}, but observed at slightly higher 
velocities. This difference could be down to the separate formation location of the lines in the expanding wind, since the periods are similar. The 
4$^{\mathrm{th}}$ panel down in Fig. \ref{fig:phase} shows the data in the power weighted average velocity channel of the significant period detected, 
($-35.0$ km s$^{-1}$) phase folded with the overall average period of $13.48$ days. (The range of significant detections of periodicity in Li\,\textsc{i} includes 
the $13.28$ and $14.27$ day periods found in H$\alpha$ and Na\,\textsc{i} D profiles respectively, so it is plausible that all these periods could be produced 
via the same mechanism.) The diagram in Fig. \ref{fig:phase} shows clear cyclic variation and when fitted with a sinusoid, shows the variations are 
roughly in-phase with the variations of the Na\,\textsc{i} D components, displayed in the panel above. The intermediate phase of both Li\,\textsc{i} and 
Na\,\textsc{i} D between that of the chromosphere, (red lines in Fig. \ref{fig:phase}) and the higher velocity blueshifted components of H$\alpha$ are 
consistent with both the presence of periodicity and the lines' formation lower down in the expanding wind.

Another metallic absorption line not as optically thick as H$\alpha$ is the Fe\,\textsc{ii} line at 5018 \AA{}, suggested by \cite{herbig66} to be 
produced further out in the wind in a `shell' feature of FU Orionis displaced $\sim -80$ km s$^{-1}$. Fig. \ref{fig:5018prof} 
shows the $\lambda$5018 line profiles, which show no indication of double peaked structure, but instead are predominately asymmetric with 
maximum absorption depth $\sim -50$ km s$^{-1}$, slightly greater than Li\,\textsc{i}, more similar to H$\alpha$ and H$\beta$. On some nights, most notably 
JD=2,454,121.333, the absorption forms a sharp dip, which could be attributed to a shell-like signature in the wind. However, this spectral feature was 
identified by \citet{calvet93} to be present when the disc rotational velocity is comparable to the expansion velocity in the wind, thus the 
`shell' signature is disingenuous. Consequently, the loss of the region of sharp decline, and the production of a more symmetric 
blueshifted absorption profile on Fig. \ref{fig:5018prof} by JD=2,454,127.377, has implications for the mass-loss rate and expansion 
velocity in the wind. The blueshifted absorption appears to extend out to $\sim -125$ km s$^{-1}$, however both the width and the strength of 
the absorption profile in general appears to vary, comparable to the H$\alpha$ and Li\,\textsc{i} $\lambda$6707 profiles respectively.

\begin{figure*}
 \centering
 \includegraphics[width=0.99\linewidth]{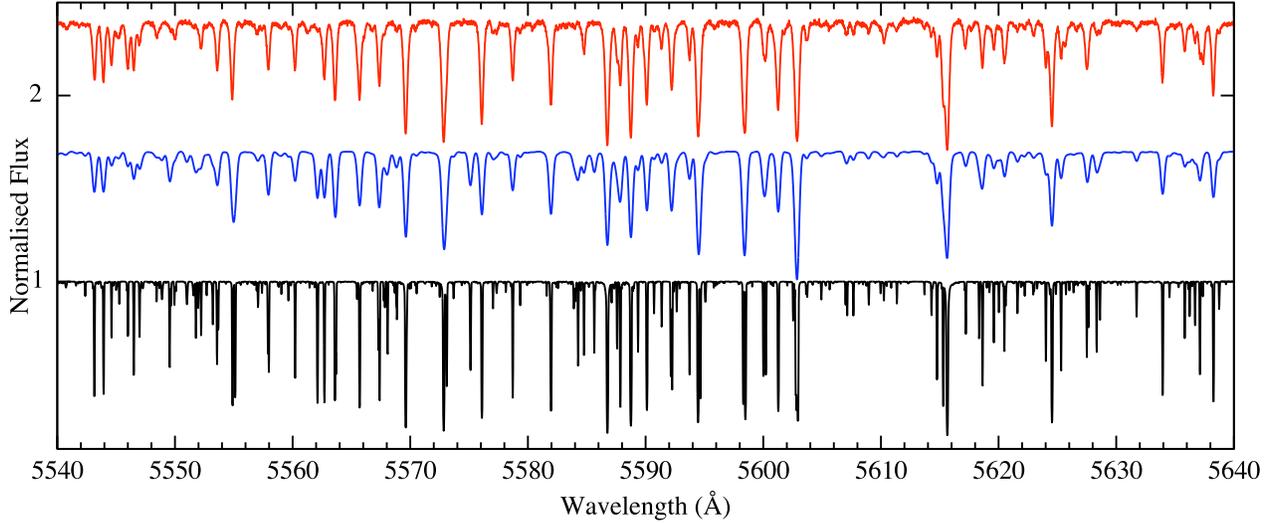}
 \caption{Reference spectra used in the cross correlations with FU Orionis, shown in the 5540--5640 \AA{} region. The bottom black line is a synthetic stellar 
spectrum of T$_{\mathrm{eff}}=6000$ K, log(g)$=3.0$, the blue line is the same spectrum smoothed with a Gaussian with FWHM of $0.37$ \AA{}, which was found to 
correspond to the FWHM of the unblended, unsaturated lines in the $\beta$ Aqr spectrum. The top red line is the spectrum of $\beta$ Aqr, 
obtained from the {\it SOPHIE} \'echelle spectrograph mounted on the $1.93$-m telescope at the OHP with a resolution of $4$ km s$^{-1}$.}
 \label{fig:5540refs}
\end{figure*}

The periodogram shown in Fig. \ref{fig:5018contour} was obtained via the method described in section \ref{sec:halphalines}, over the same 
parameter space for comparison. It shows less significant detections of a period than the Li\,\textsc{i} $\lambda$6707 profile, but there are 
still periods significant to the $0.01$ FAP with velocities from $-43.5$ to $-24.0$ km s$^{-1}$ with values between $12.31$ and $15.68$ days. 
This corresponds to the largest feature in Fig. \ref{fig:5018contour}, which has a power weighted average period of $13.57$ days. 
The dominant periodic feature detected in this line profile is yet again of a similar value to that of H$\alpha$, H$\beta$ Na\,\textsc{i} and Li\,\textsc{i}, 
providing further evidence that there is a wind period in FU Orionis. This is confirmed in Fig. \ref{fig:phase}, where the 4$^{\mathrm{th}}$ panel from the 
bottom shows the Fe\,\textsc{ii} $\lambda$5018 data in the power weighted average velocity of the periodic feature ($-39.0$ km s$^{-1}$), phase 
folded with the average overall wind period of $13.48$ days. The cyclic variations are fitted with a sinusoid, (solid blue line), and the fit is 
better than that for Na\,\textsc{i} D and Li\,\textsc{i} in the two panels above. However, unlike the previous line profiles, the phase of the Fe\,\textsc{ii} 
line does not lie within the range expected for formation in the expanding wind, and seems to respond even before the chromospheric lines in the top two 
panels in Fig. \ref{fig:phase}. Although there are less significant detections of a period to less than $0.01$ FAP visible in Fig. \ref{fig:5018contour} 
than in Li\,\textsc{i} $\lambda$6707 (see Fig. \ref{fig:licontour}), possibly due to the high temperature sensitivity of the Li\,\textsc{i} line 
\citep{calvet93}, the amplitude of variation is greater in the Fe\,\textsc{ii} line at $\lambda$5018 (see Fig. \ref{fig:phase}). On examination of Fig. 
\ref{fig:5018prof}, the velocities at which this period was detected, (shown by the shaded region) seem to correlate with the appearance and disappearance 
of the absorption `dip', confirming the assumption of the period's origination in a wind. The fact that there are less significant detections of 
periodicity could indicate that the effects of the mechanism for producing this period in the wind may become reduced at greater distances 
from the disc where the Fe\,\textsc{ii} $\lambda$5018 absorption line is thought to originate.

As in the Li\,\textsc{i} $\lambda$6707 profile, there is also another feature visible on the periodogram in Fig. \ref{fig:5018contour} at slightly 
lower frequency and more negative velocity than the dominant feature, corresponding to a weak detection of periodicity, above the $0.01$ FAP level, 
(power \textgreater{} $6.73$). The range in velocity of this yellow feature is $-128.0$ to $-123.5$ km s$^{-1}$ and spans periodicities between $19.52$ to 
$23.91$ days. This is comparable to the detection discussed in Li\,\textsc{i} $\lambda$6707, where the authenticity of the detection was unclear. 
In Fe\,\textsc{ii} $\lambda$5018 this detection has a power weighted period of $21.66$ days, slightly lower than the 
comparable signal in Li\,\textsc{i} $\lambda$6707, yet there is a substantial overlap in the significant periods detected in both lines. 
The velocities at which the period manifests itself are slightly higher in Fe\,\textsc{ii} $\lambda$5018 than in Li\,\textsc{i} $\lambda$6707, however 
this could be ascribed to the different in formation locations in an expanding wind. The recurrence of this feature at a comparable periodicity 
to an ambiguous detection in Li\,\textsc{i} $\lambda$6707 that affects a larger area of surrounding parameter space in Fig. \ref{fig:5018contour} 
then in \ref{fig:licontour}, makes this detection slightly less likely due to noisy line profiles and may in fact be linked to the main source of 
periodicity in both cases.

\section{Cross Correlations}
\label{sec:ccf}

\begin{figure*}
 \centering
 \includegraphics[angle=-90,width=0.99\linewidth]{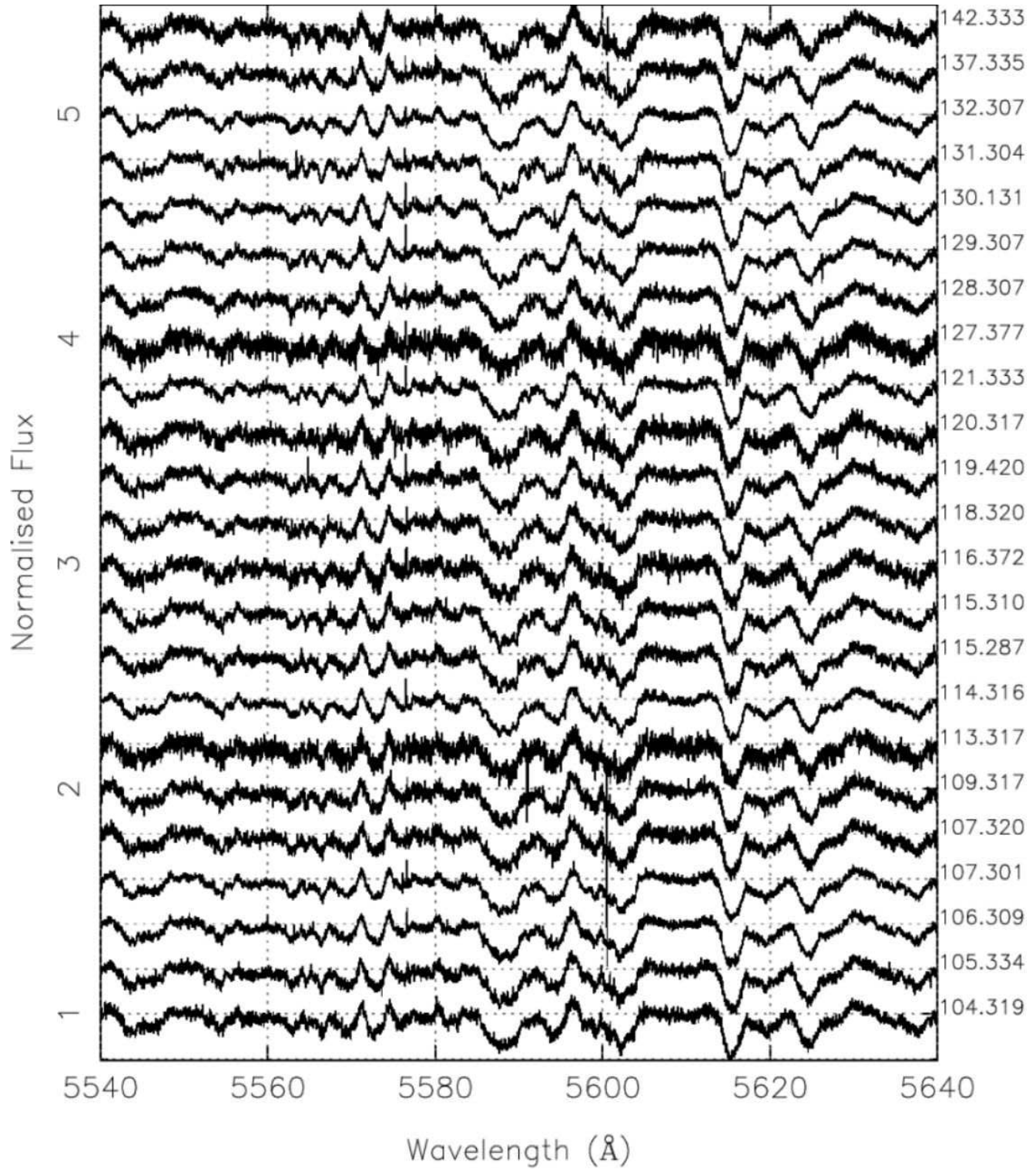}
 \caption{Spectra of FU Orionis in the 5540--5640 \AA{} region obtained from the {\it SOPHIE} \'echelle spectrograph mounted on the $1.93$-m telescope at the OHP.
The $21$ spectra have a resolution of $4$ km s$^{\mathrm{-1}}$ and cover a time frame of $38$ nights, with the Julian dates of the spectra -2,454,000 
displayed on the right hand y-axis. All spectra have been barycentrically corrected, normalised to unity, placed in the rest frame of FU Orionis, continuum 
subtracted with a bias level of $0.97$ and shifted to show night to night variability.}
 \label{fig:5540spectra}
\end{figure*}

\begin{figure*}
 \centering
 \begin{minipage}{0.45\linewidth}
  \centering \includegraphics[width=0.99\linewidth]{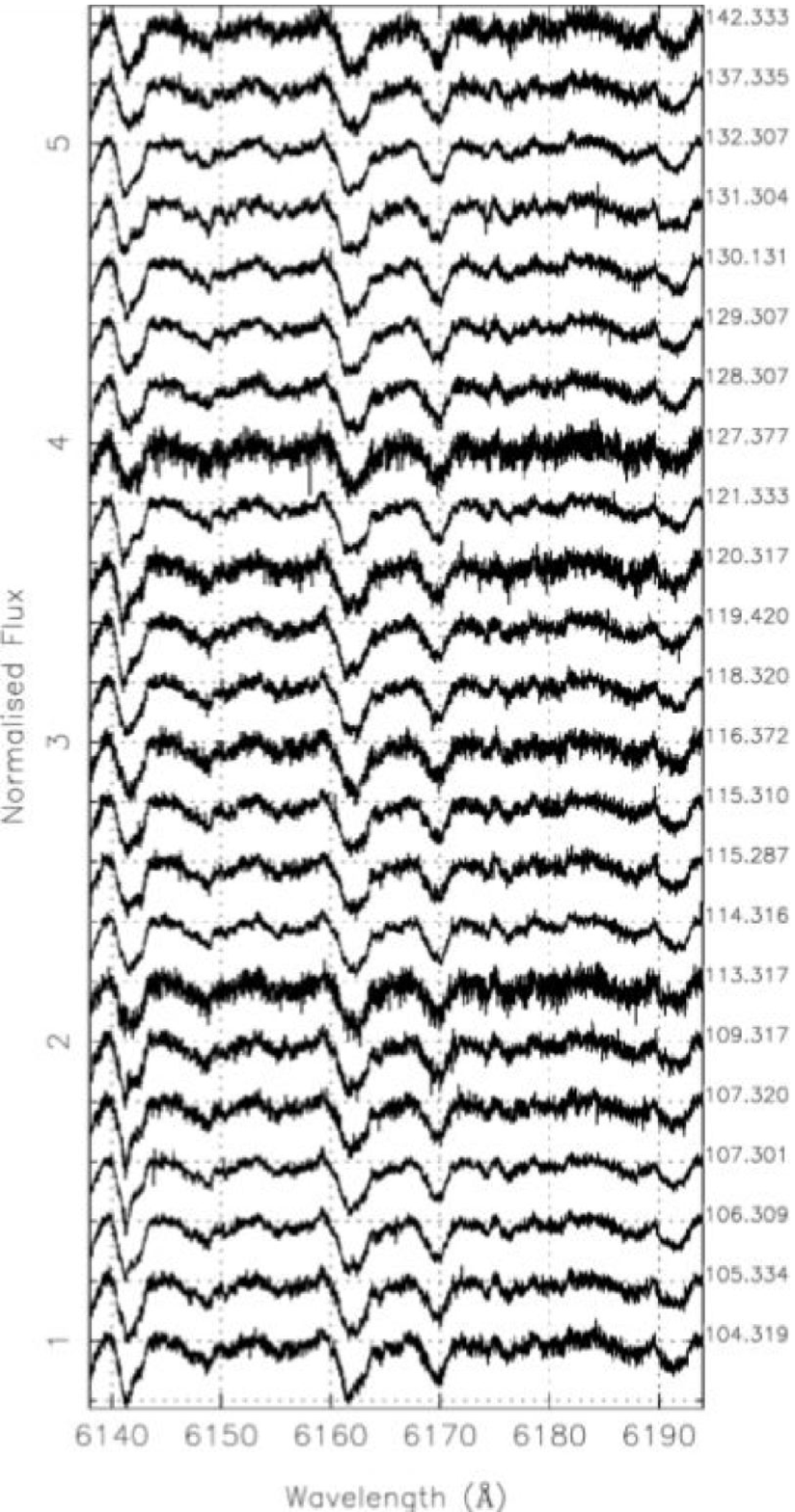}
 \caption{Same as Fig. \ref{fig:5540spectra}, but for the 6170 \AA{} region using a bias level of $0.98$ for the continuum subtraction.}
 \label{fig:6138spectra}
\end{minipage}
\hspace{1cm}
\begin{minipage}{0.45\linewidth}
 \centering
 \includegraphics[width=0.99\linewidth]{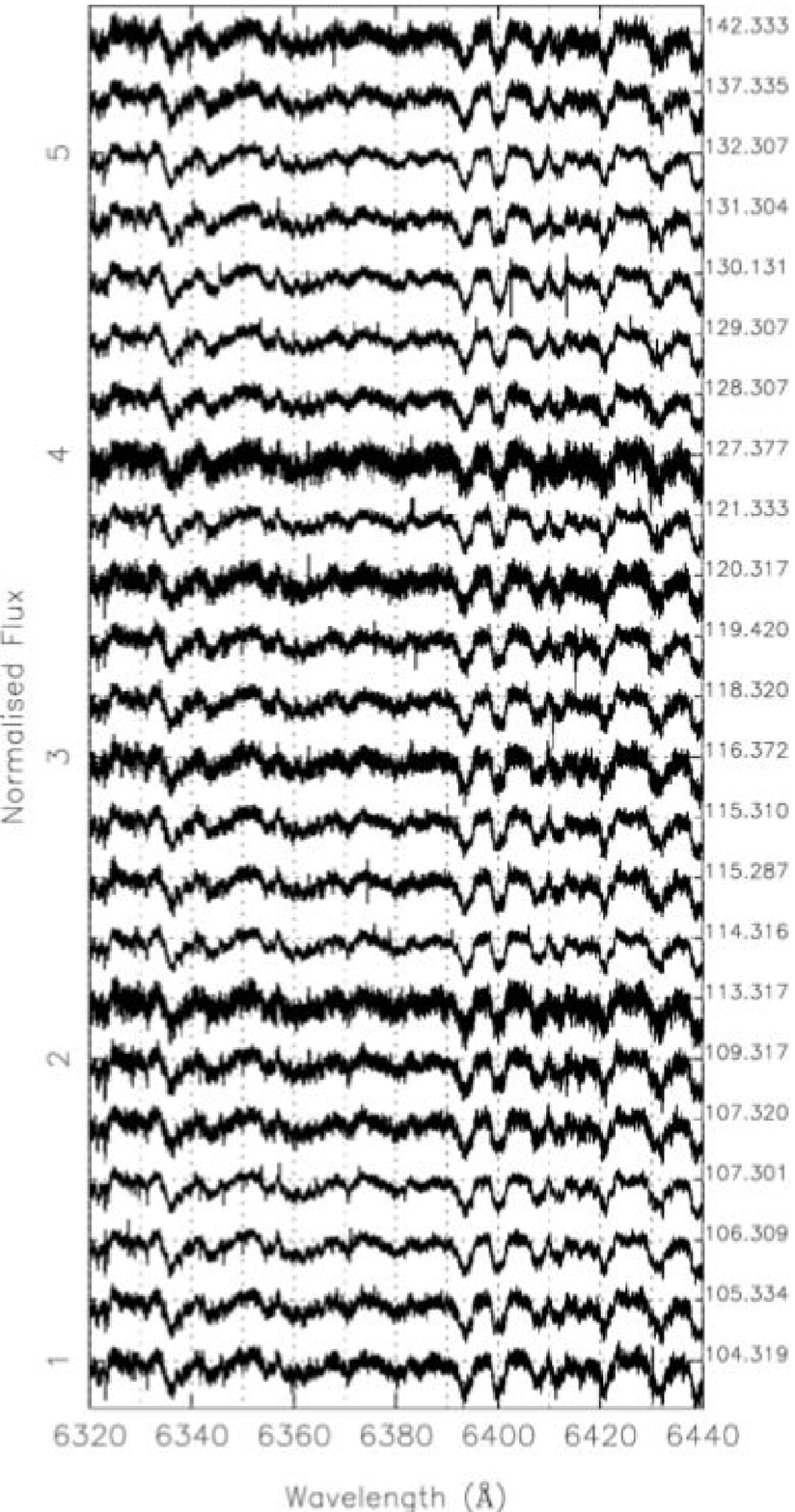}
 \caption{Same as Fig. \ref{fig:5540spectra}, but for the 6320--6440 \AA{} region using a bias level of $0.98$ for the continuum subtraction.}
 \label{fig:6320spectra}
\end{minipage}
\end{figure*}

\begin{figure*}
\begin{minipage}{0.99\linewidth}
 \centering
 \includegraphics[width=0.75\linewidth]{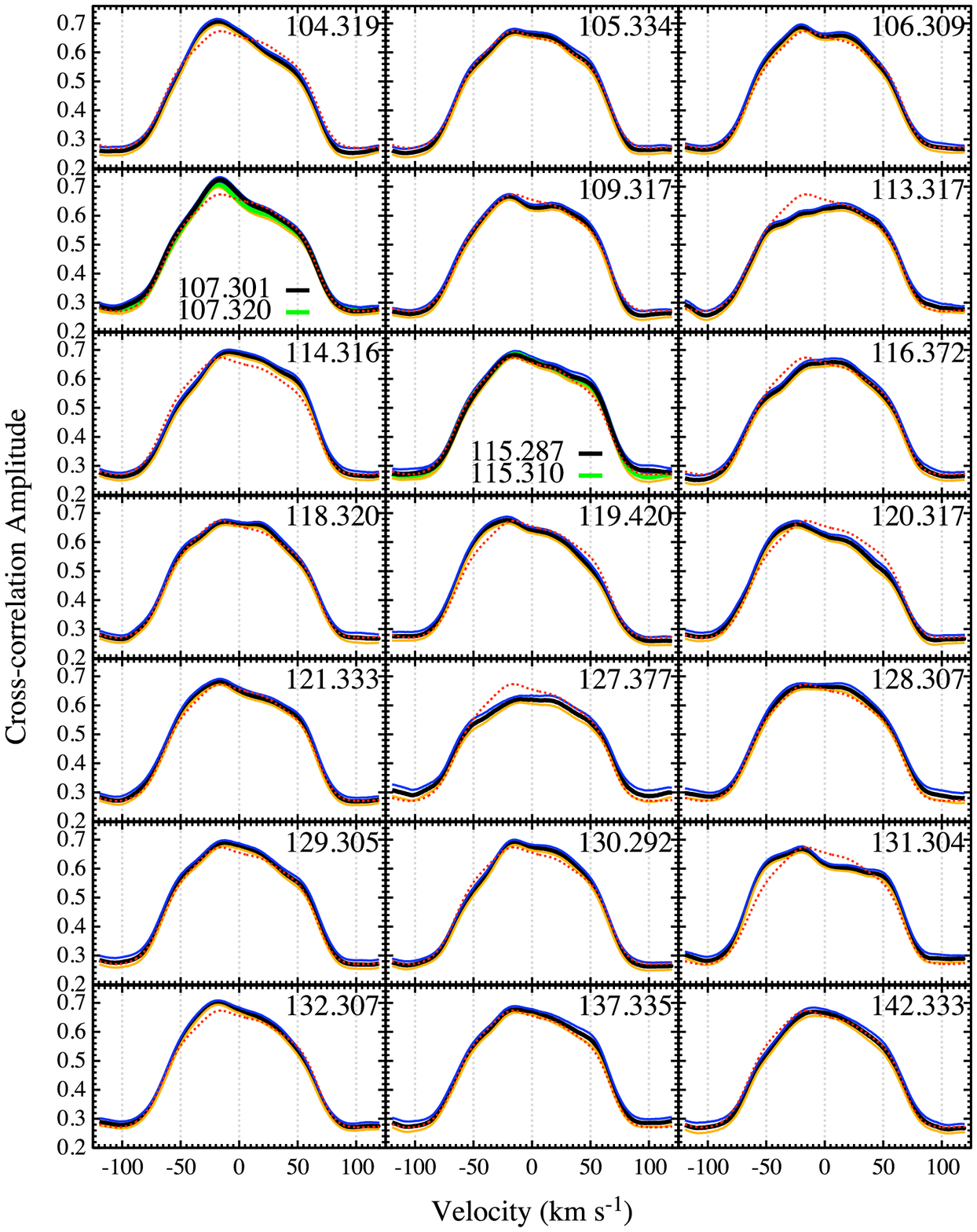}
 \caption{Cross correlations of FU Orionis in the 6138--6194 \AA{} region with a reference spectrum of $\beta$ Aqr obtained from the {\it SOPHIE} \'echelle 
spectrograph mounted on the $1.93$-m telescope at the OHP. The cross correlation function is the black line, with the upper and lower $3\sigma$ errors 
shown by the dark blue and light blue lines respectively. When two observations were taken on a single night a dark green line is used to represent 
the second observation. The average over all cross correlations in this region is shown by the red dashed line.}
 \label{fig:6138ccf}
\end{minipage}
\begin{minipage}{0.99\linewidth}
 \centering
\hspace{1cm}
 \includegraphics[width=0.55\linewidth]{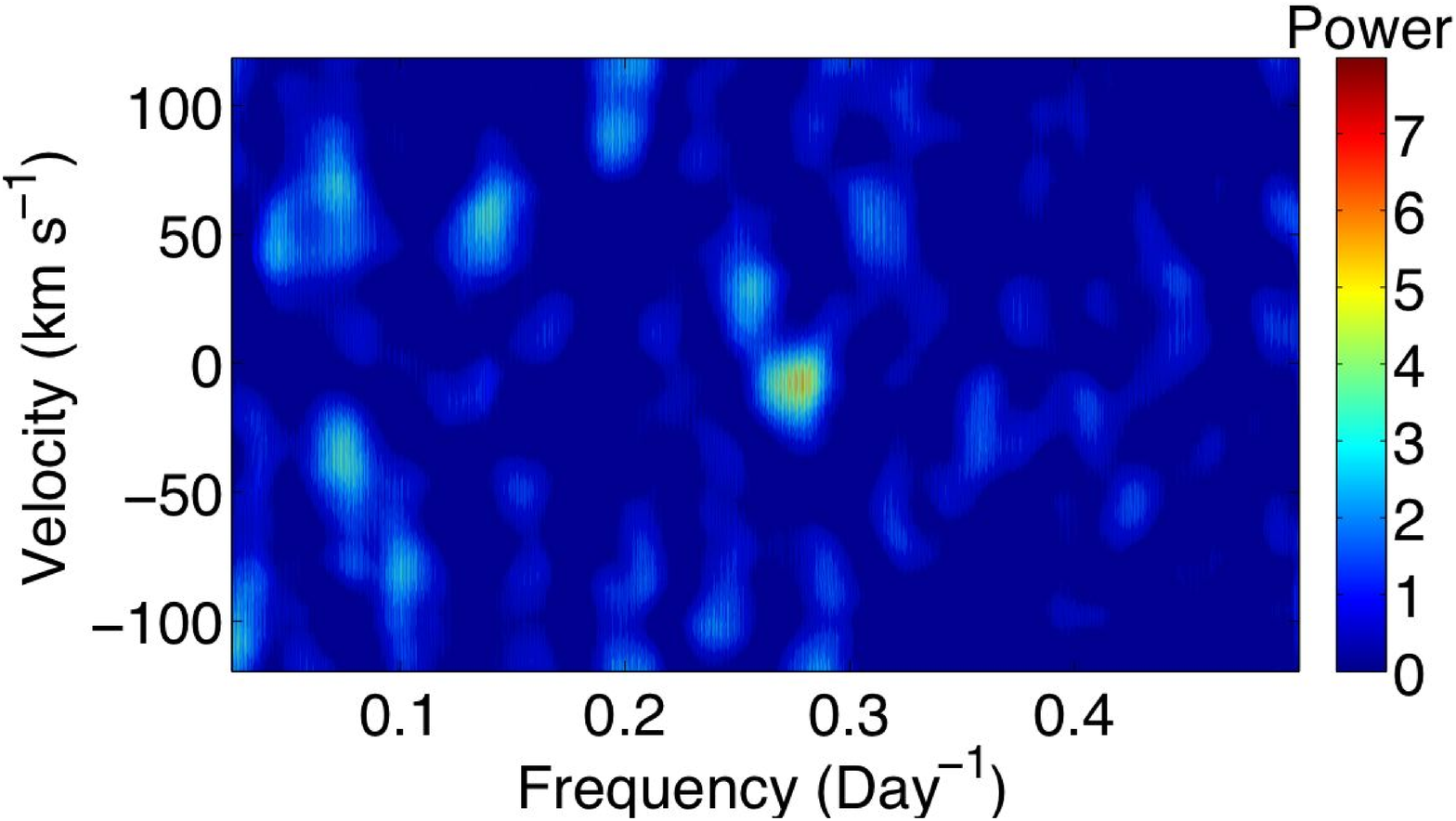}
 \caption{Contour periodogram showing the significant periods found in the 6138--6194 \AA{} region. The $0.01$ FAP corresponds to power \textgreater{} $6.88$.}
 \label{fig:6138contour}
\end{minipage}
\end{figure*}

\begin{figure*}
\begin{minipage}{0.99\linewidth}
  \centering
 \includegraphics[width=0.75\linewidth]{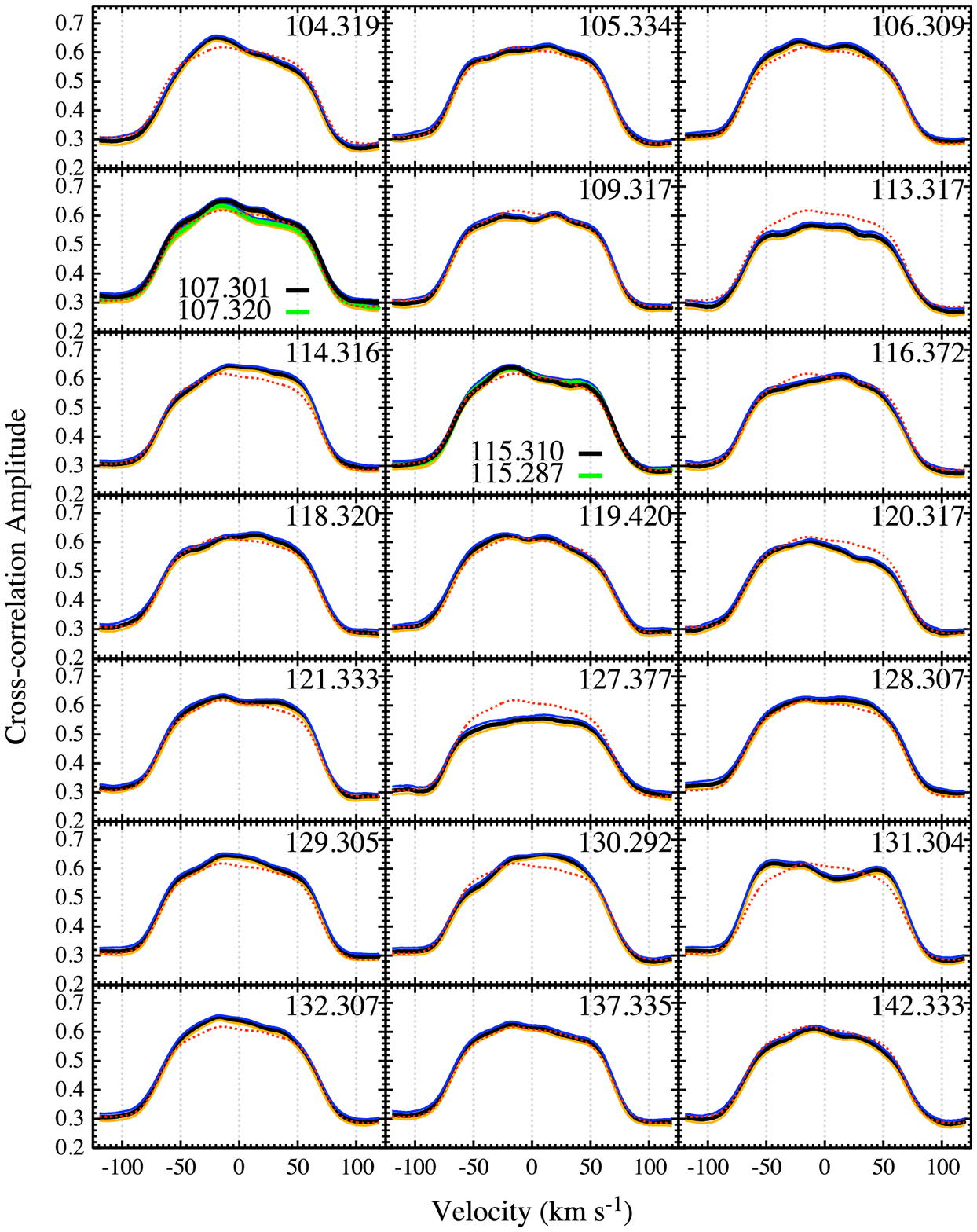}
 \caption{Same as Fig. \ref{fig:6138ccf}, but for the 6320--6440 \AA{} window.}
 \label{fig:6320ccf}
\end{minipage}
\begin{minipage}{0.99\linewidth}
 \centering
\hspace{1cm}
 \includegraphics[width=0.55\linewidth]{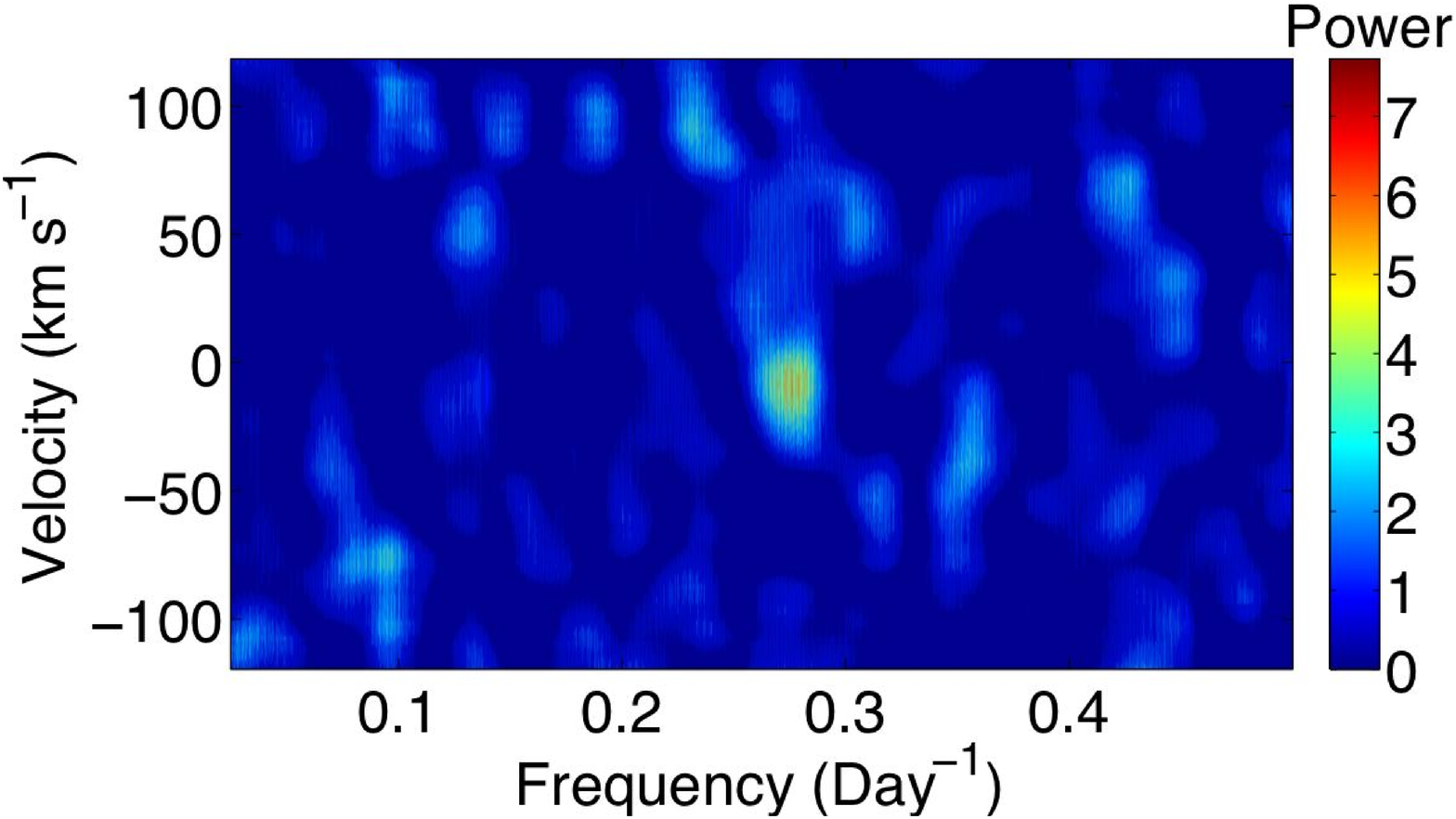}
 \caption{Contour periodogram showing the significant periods found in the 6320--6440 \AA{} region. The $0.01$ FAP corresponds to power \textgreater{} $6.73$}
 \label{fig:6320contour}
\end{minipage}
\end{figure*}

\begin{figure*}
\begin{minipage}{0.99\linewidth}
  \centering
 \includegraphics[width=0.75\linewidth]{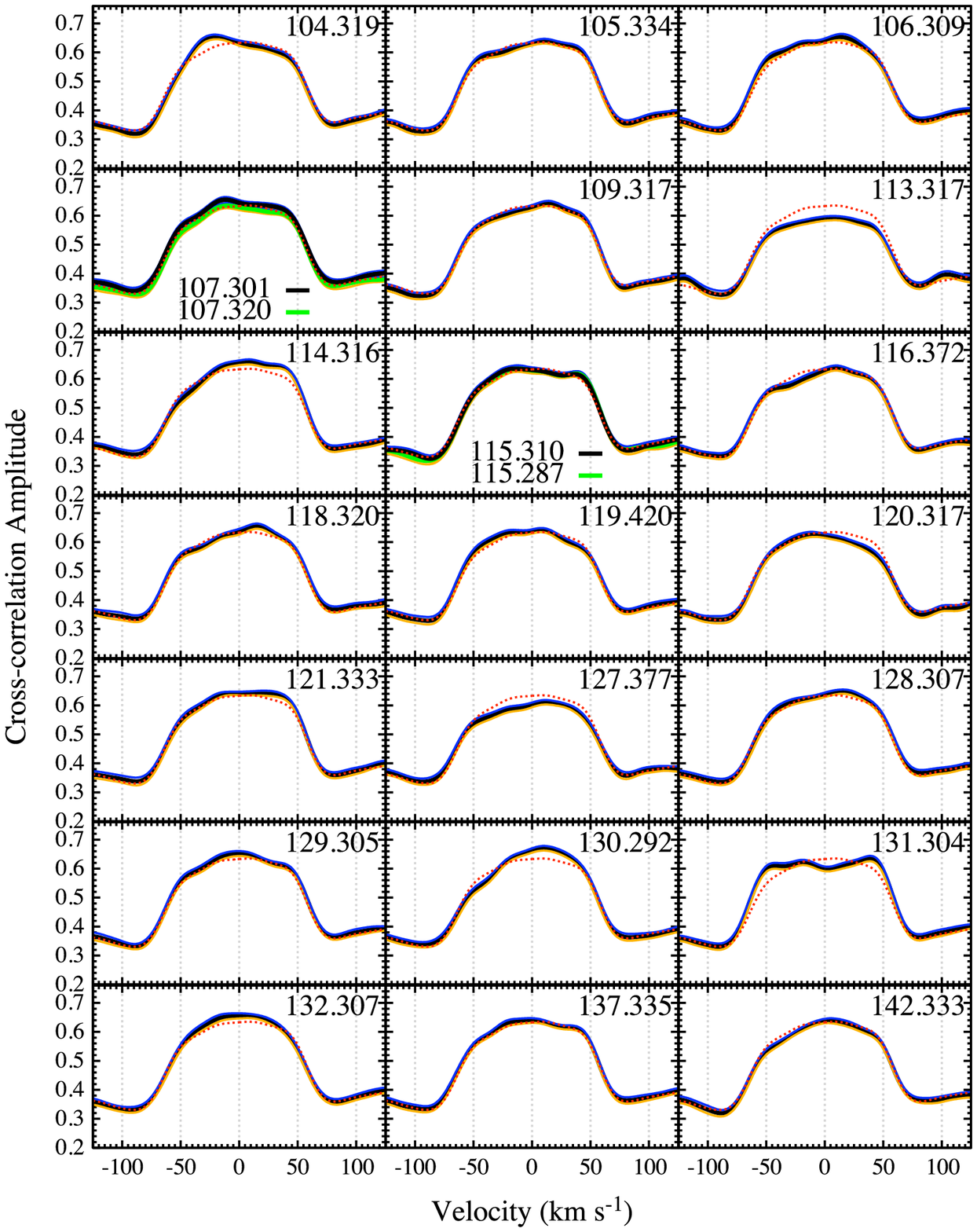}
 \caption{Same as Fig. \ref{fig:6138ccf}, but for the 5540--5640 \AA{} window, which was continuum subtracted with a lower bias of $0.97$ and cross 
correlated with a G0 synthetic stellar template at T$_{\mathrm{eff}}=6000$ K, log(g)$=3.0$.}
 \label{fig:5540ccf}
\end{minipage}

\begin{minipage}{0.46\linewidth}
\centering
\hspace{1cm}
\includegraphics[width=0.99\textwidth]{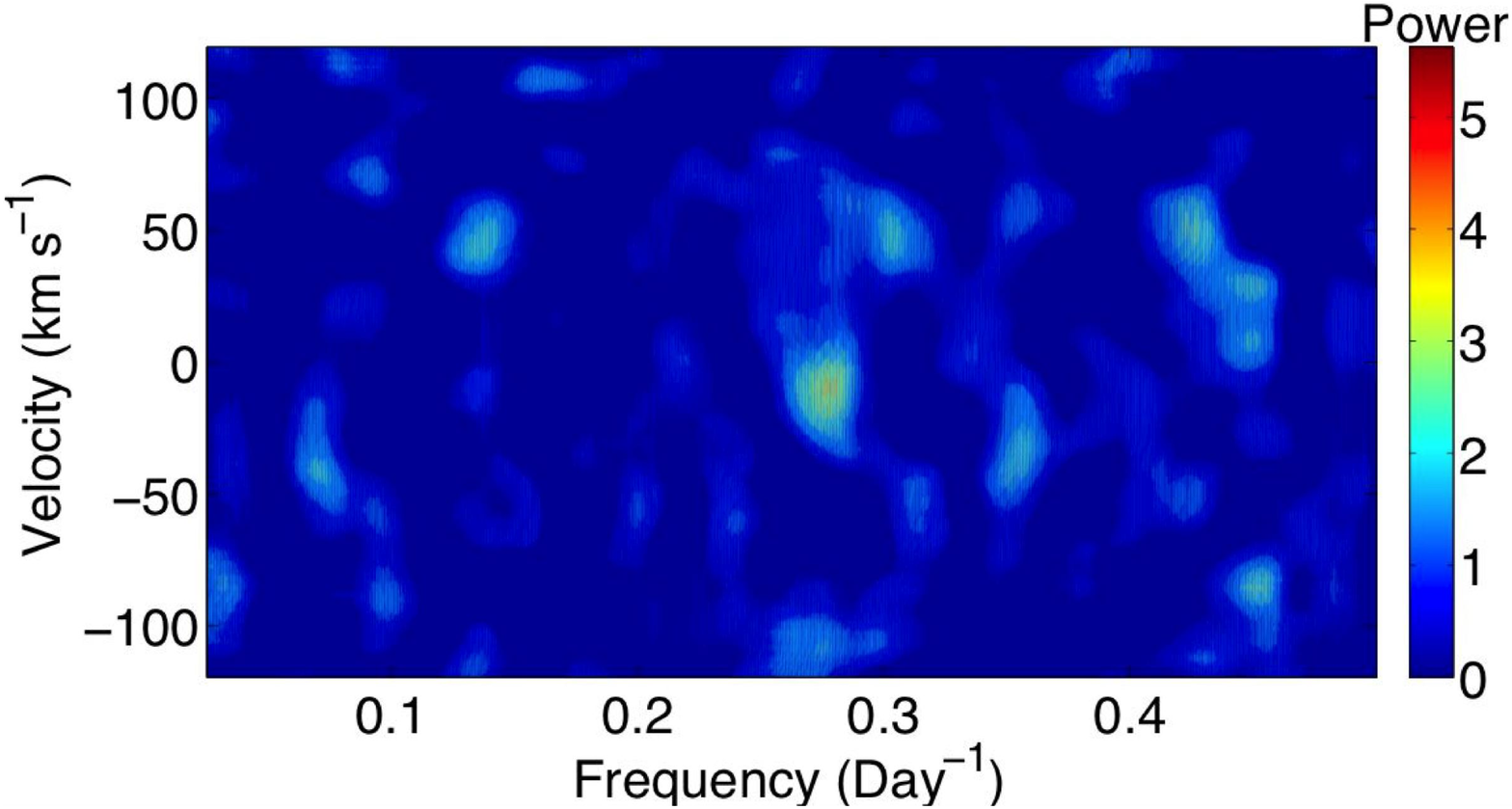}
\caption{Contour periodogram of the 5540--5640 \AA{} cross correlation, which was continuum subtracted using a lower bias of $0.97$ and cross 
correlated with observed reference template of $\beta$ Aqr. There are no detections in the cross correlation function below the $0.01$ FAP which corresponds to 
power \textgreater{} $5.88$.}
\label{fig:5540contour}
\end{minipage}
\hspace{1cm}
\begin{minipage}{0.46\linewidth}
\centering
\hspace{1cm}
\includegraphics[width=0.99\textwidth]{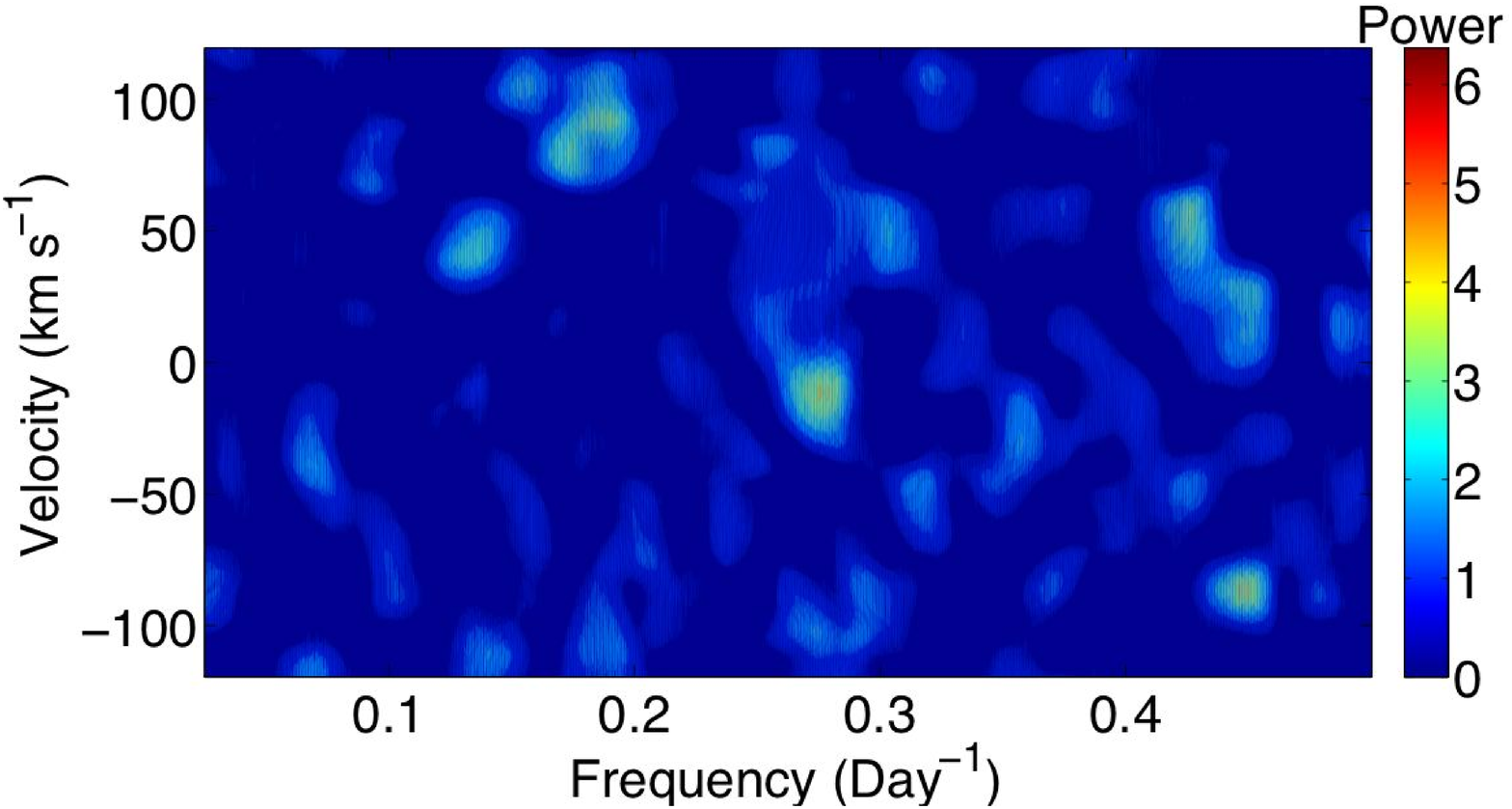}
\caption{Contour periodogram showing the significant periods found in the 5540--5640 \AA{} cross correlation, which was continuum subtracted with a 
lower bias level of $0.97$ and cross correlated with a G0 synthetic stellar template at T$_{\mathrm{eff}}=6000$ K, log(g)$=3.0$. 
The $0.01$ FAP corresponds to power \textgreater{} $6.16$.}
\label{fig:5540contoursynthetic}
\end{minipage}
\end{figure*}

Some of the metal lines were too weak and too noisy to analyse the line profile alone. In order to increase the signal-to-noise ratio, these 
profiles were cross correlated with a template spectrum of $\beta$ Aqr and a synthetic stellar spectrum of a G0 star with T$_{\mathrm{eff}}=6000$ K, 
log(g)$=3.0$  to identify any potential effects of relative line shifts, as described in section \ref{sec:obs}. The standard spectra in the 5540--5640 \AA{}
spectral region are shown in Fig. \ref{fig:5540refs}, where the bottom black line is the synthetic spectrum and the blue line shows the Gaussian smoothed 
synthetic spectrum to match the FWHM of the unblended, unsaturated lines in the observed spectrum of $\beta$ Aqr, shown by the top red line. 
The regions cross correlated were chosen in order to compare with previous results by \citet{herbig03} where a $3.542$ day period has
already been identified in the centroid of the cross correlation, and \citet{hartmann87a}, who found variability in their cross correlation
profile of the 6170 \AA{} region. The regions investigated by \citet{herbig03} were the 6320--6440 \AA{} region and the 5540--5640 \AA{} 
region, which were initially identified in order to avoid lines with wind components. The normalised {\it SOPHIE} \'{e}chelle spectra 
taken from 2007 January 3$^{\mathrm{rd}}$ to February 10$^{\mathrm{th}}$ are shown in Figs. \ref{fig:5540spectra}, \ref{fig:6138spectra} and
\ref{fig:6320spectra}. The spectra in the 5540--5640 \AA{} region, (see Fig. \ref{fig:5540spectra}) were normalised using a lower bias level ($0.97$ 
instead of $0.98$), because the continuum appeared under-subtracted when the higher value was used. All spectra show evidence of weak absorption lines and 
some weak lines appear doubled. This is most notable in Fig. \ref{fig:6138spectra} where the 6162.17 \AA{} Ca\,\textsc{i} and 6169.56 \AA{} Ca\,\textsc{i} 
absorptions appear doubled on some nights, indicating a disc origin, although the profiles are highly variable. Other lines noted by \citet{hartmann87a} 
to show evidence of double peaked structure in this spectral region are the 6147.8 \AA{} Fe\,\textsc{i} and Fe\,\textsc{ii} line blend, 6151.62 \AA{} 
Fe\,\textsc{i}, 6157.73 \AA{} Fe\,\textsc{i}, and 6166.44 \AA{} Ca\,\textsc{i}. There does appear to be some evidence of a skewed absorption profile in the 
6141.73 \AA{} Fe\,\textsc{i} line, indicating the presence of mass-loss. Cross correlating these spectral regions provides a distinct advantage as not only 
does it produce an average profile weighted according to the standard line strengths, but it also allows for lines which overlap and blending, abundant in 
these regions, which is clearly a problem for identifying the weak lines in Figs. \ref{fig:5540spectra}, \ref{fig:6138spectra} and \ref{fig:6320spectra}. 
The cross correlation functions (CCFs) of each night were tested for periodicity at 5000 different frequencies between $2$--$40$ days to quantitatively
compare the results with the line profiles in sections \ref{sec:balmerlines}, \ref{sec:NaD} and \ref{sec:metallic}.

The CCF of the 6138--6194 \AA{} region, displayed in Fig. \ref{fig:6138ccf}, showed the clearest detection of a period significant to less than 
$0.01$ FAP, (power \textgreater{} $6.88$) in velocity channels $-14.0$ to $-1.5$ km s$^{-1}$. The significant periods range from $3.52$ to $3.70$ days and 
have a power weighted average of $3.61$ days. This region is also clearly visible above the noise in Fig. \ref{fig:6138contour}. This period is similar 
in value to that detected previously, yet it is distinctly different in signature, corresponding to a cyclic strengthening/weakening of the blue wing of 
the line profiles in this region. To compare with \cite{herbig03}, the radial velocity of centroid of the CCF was found for each observation 
and tested for periodicity between $2$ and $40$ days, yet no significant period was found below the $0.01$ FAP. This was unexpected, as the periodicity has 
only been observed on the blue wing of the CCF. To make a more direct comparison, the spectra were degraded to a resolution of $13$ km s$^{-1}$ 
used by \citet{herbig03}. At this resolution a longer period of $13.12$ days, (comparable to the periodicity previously observed in the line profiles directly) 
was detected in the centroid, (or weighted average velocity) of the cross correlation in the 6138--6194 \AA{} region, which was significant to the $0.01$ FAP 
level. This could possibly indicate the effects of the wind are more significant in this spectral region, than in the regions investigated by \citet{herbig03}, 
and the shorter periodic effects may be masked in the centroids due to the other variable features in the CCF, most notably the effects of the wind. The power 
weighted average radial velocity of the `blob' in Fig. \ref{fig:6138contour} was $-7.5$ km s$^{-1}$, and the phase folded data at this velocity is shown in the 
2$^{\mathrm{nd}}$ panel from the bottom in Fig. \ref{fig:phase}. The precise velocity at which this 
periodicity is detected is difficult to validate since spectral lines in the observed G0 Ib template of $\beta$ Aqr may be shifted due to nightly 
fluctuations in air mass, which could ultimately shift the velocity peaks in the cross correlations. However, the velocity range of the periodicity 
was reinforced by cross correlations with the synthetic stellar spectrum at T$_{\mathrm{eff}}=6000$ K log(g)$=3.0$, which confirmed periodicity between $3.54$ 
to $3.71$ days to below $0.01$ FAP spanning velocities from $-16.0$ to $-4.0$ km s$^{-1}$, with a power weighted average period of $3.62$ days.

The detections below the $0.01$ FAP level, (power \textgreater{} $6.73$) in the 6320--6440 \AA{} region cross correlated with the observed template of $\beta$ Aqr 
were periods from $3.56$ to $3.71$ days, spanning velocity channels from $-18.5$ to $0.0$ km s$^{-1}$, which produced a weighted average period of $3.63$ days. 
Fig. \ref{fig:6320ccf} shows the resulting CCFs, which look very similar to the cross correlations of the 6138--6194 \AA{} region in Fig. \ref{fig:6138ccf}. 
Fig. \ref{fig:6320contour} shows the significant detections with the red blob at the centre of the figure, at a frequency of $0.28$ days$^{-1}$. When the 
centroids of the CCFs in this region were tested for periodicity, yet again there was none found to be significant to below the $0.01$ FAP level. However, 
when the spectra and observed template were degraded to the resolution used by \citet{herbig03} in this region, the centroids of the cross 
correlation peaks were found to have a periodicity of $13.69$ days significant to the $0.01$ FAP level. This could possibly indicate an increase in the mass-loss 
rate, which would manifest wind signitures more readily in this epoch. This may also be a possible explanation why 
\citet{herbig03} noted a poor fit to the sinusoid for the observations taken in the year 2000. The velocities at which this periodic signal was detected were 
confirmed by the cross correlations with the synthetic stellar spectrum, which also detected a period of $3.63$ days between $-19.0$ to $-1.0$ km s$^{-1}$, 
significant to below the $0.01$ FAP. The phase folded data at $-9.0$ km s$^{-1}$ fitted with a sinusoid shown in the blue solid line, can be seen on the 
3$^{\mathrm{rd}}$ panel from the bottom in Fig. \ref{fig:phase} with error bars plotted to the $3$ sigma level.

Our analysis showed no significant periodicity below the $0.01$ FAP corresponding to a power of $5.88$ in the 5540--5640 \AA{} region when cross-correlated 
with $\beta$ Aqr. Unsurprisingly, no period was detected in the centroid of the cross correlation of the spectra and observed template degraded to the same 
resolution as \citet{herbig03} in this spectral region. However, a period between $3.63$--$3.65$ days was found in the same spectral region when cross correlated 
with the synthetic template, which was significant below the $0.01$ FAP, corresponding to powers above $6.16$ in this case. The weighted average period was found 
to be $3.64$ days, from $-15.0$ to $-9.0$ km s$^{-1}$ and centred at $-12.0$ km s$^{-1}$. The resulting cross correlations with the smoothed synthetic 
spectrum are shown in Fig. \ref{fig:5540ccf} with the periodograms displaying the null detection using the $\beta$ Aqr template, and the significant periods found 
with the synthetic template in Figs. \ref{fig:5540contour} and \ref{fig:5540contoursynthetic} respectively. The background level at the edges of the CCFs are 
slightly higher in Fig. \ref{fig:5540ccf}, compared to Figs. \ref{fig:6320ccf} and 
\ref{fig:6138ccf}, because of the lower bias level used in the continuum subtraction. However the shape remains unaffected in this adjustment, and on 
visual examination the variability seems to correlate with the cross correlations of the 6320--6440 \AA{} and 6138--6194 \AA{} spectral regions using the observed 
template of $\beta$ Aqr. The absence of periodicity in the cross correlations of the 5540--5640 \AA{} region with the observed template and yet the 
significant detection with the synthetic template, could be manifested by the slightly different relations between spectral line strengths in these two 
spectra, shown with the top two lines plotted in Fig. \ref{fig:5540refs}. Figs. \ref{fig:5540contour} and \ref{fig:5540contoursynthetic} are extremely 
similar and both show a clear feature around the periodicity detected to be significant to below a FAP of $0.01$ in the cross correlations with the synthetic 
template. 
Therefore although this feature is not detected to be significant in Fig. \ref{fig:5540contour} to such a high level when the observed template of $\beta$ 
Aqr is used, it is still a clear feature above the noise in the periodogram and hence the lack of high significance of the feature in this case may be due to the 
slight differences in the line weightings producing slightly different weights in the cross correlation function and consequently, the periodogram. In the case of 
the synthetic standard, these slight differences have contributed to the significance of the cyclic variability in this spectral region. The absence of any 
significance to less than $0.01$ FAP with the $\beta$ Aqr standard may therefore not indicate the lack of cyclic variability, but rather inadequate signal 
strength in this spectral region possibly due to a different mix/ratio of line species in the observed template or a slightly different origin of these 
absorption lines within the disc itself in comparison to the spectral regions previously discussed. The bottom panel of 
Fig. \ref{fig:phase} shows the data at $-12.0$ km s$^{-1}$ phase folded with a period of $3.64$ days and fitted with a sinusoid. For the reasons discussed 
above, the amplitude of variation is small in comparison with the cross correlations in the other spectral regions, shown in the 2 panels above in 
Fig. \ref{fig:phase}.

The in-phase nature of the periods found in the cross correlation functions, along with the extensive overlap in velocity and range of significant periods 
identified to less than $0.01$ FAP is strong evidence that all these cyclic variations are produced from a single origin. Consequently, the three periods 
identified above were combined to find a variance weighted total average period of $3.64$ days, which was used to phase fold the data in all regions plotted 
in Fig. \ref{fig:phase}. The periodicity could be attributed to a slow moving wind, as it is predominantly present at small negative velocities. However 
it does not appear to be related to the wind periods found in the line profiles in sections \ref{sec:balmerlines}, \ref{sec:NaD} and \ref{sec:metallic}, as 
it is roughly one quarter of the value.

To complete the analysis of the cross correlations, the FU Orionis spectra from each night were cross correlated with the average spectrum and the 
resulting CCFs were tested for periodicity via the same analysis. Interestingly in the 6138--6194 \AA{} region, this resulted in periods between $12.31$ to 
$14.32$ days, significant to below $0.01$ FAP with a power weighted average of $13.19$ days, which are similar to the periodicities detected in the 
centriods of the spectra degraded to the resolution of \citet{herbig03} and in sections \ref{sec:balmerlines}, \ref{sec:NaD} and 
\ref{sec:metallic}. Additionally, this periodicity was detected at velocities between $-80$ to $-24$ km s$^{-1}$, more comparable to the fast velocities of 
an expanding wind. No such periodicity was detected in the other spectral regions, possibly due to a different mix of line species and their selection by 
\citet{herbig03} to avoid line with wind components. There was no evidence of the $3.64$ day period in this analysis. However, the resulting cross correlations 
are extremely broad due to the large FWHM of spectral lines in the FU Orionis spectra in comparison with the FWHM of the standard templates, (see Figs. 
\ref{fig:5540refs} and \ref{fig:5540spectra} for comparison). Therefore the CCFs produced using the average spectra as a template are insensitive to 
small changes around the line centre, which would be necessary to detect the $3.64$ day period. Consequently, there appears to be more than one mechanism 
producing asymmetries in FU Orionis.

\section{Discussion}
\label{sec:dicus}

\begin{table}
\begin{minipage}{0.99\linewidth}
\centering
  \caption{Summary of periodicities detected - (1) Line/spectral region where periodicities were detected; 
(2) Weighted average velocity of periodic feature with the limits showing the full range of velocity channels 
where periodicity was significant to below the $0.01$ FAP; (3) Weighted average periodicity detected with the full range
of periodicities detected above the $0.01$ FAP shown in the limits.}
\label{tab:results}
  \begin{tabular}{lcc} \hline
\hline
(1)&(2)&(3)\\
Spectral Region&Velocity&Period \\
&(km s$^{-1}$)&(Days) \\
\hline\smallskip
H$\alpha$	   	     & 10.0$^{+75.5}_{-22.5}$ & 13.27$^{+3.02}_{-1.68}$\\\smallskip
		             & -109.5$^{+4.0}_{-12.5}$ & 13.85$^{+0.99}_{-0.72}$\\\smallskip
H$\beta$ 	    	     &    3.5$^{+7.5}_{-11.5}$ & 12.89$^{+1.79}_{-4.07}$\\\smallskip
Na\,\textsc{i} D$_{1}$        &   -10.0$^{+4.5}_{-6.0}$ & 14.49$^{+3.38}_{-2.08}$\\\smallskip
Na\,\textsc{i} D$_{2}$	     &  -12.0$^{+6.5}_{-13.0}$ & 14.09$^{+3.46}_{-2.50}$\\\smallskip
Fe\,\textsc{ii} $\lambda$5018 &  -39.0$^{+15.0}_{-4.5}$ &13.57$^{+2.11}_{-1.26}$\\\smallskip
Li\,\textsc{i} $\lambda$6707  &   -35.0$^{+8.0}_{-8.0}$ & 14.81$^{+4.97}_{-2.60}$\\
\hline\smallskip
6138--6194\AA{}$^{a}$  &-7.5$^{+6.0}_{-6.5}$&3.61$^{+0.09}_{-0.09}$\\\smallskip
6320--6440\AA{}$^{a}$  &-9.0$^{+9.0}_{-9.5}$&3.63$^{+0.08}_{-0.08}$\\\smallskip
5540--5640\AA{}$^{b}$  &-12.0$^{+3.0}_{-3.0}$&3.64$^{+0.01}_{-0.02}$\\
\hline
  \end{tabular}
\medskip
\end{minipage}
$^{a}$ The periodicities in these regions were detected in the cross correlation function using 
an observed template of a G0Ib star namely $\beta$ Aqr.
$^{b}$ The periodicity in this region was detected in the cross correlation function
using a synthetic spectrum of a G0 star with T$_{\mathrm{eff}}=6000$ K, log(g)$=3.0$ as a reference template.
\end{table}

A summary of the velocity channels where periodicity was detected to be significant below the $0.01$ FAP level in each spectral region is shown in Table 
\ref{tab:results}. Our high resolution spectral monitoring of FU Orionis has yielded two principle results:

i) We have confirmed the periodic variability of blue shifted H$\alpha$ absorption claimed by \citet{herbig03}. The `centre of power' for periodic 
variation is at a period of $13.48$ days, (which we will henceforth term `the wind period') but the range of periods suggested by our analysis 
(see Table \ref{tab:results}), also includes the $14.847$ day period found by \citet{herbig03}. In addition, we have for the first time, found periodic variation 
(at the same period) in the redshifted emission component of H$\alpha$, the redshifted absorption component of H$\beta$, and the blueshifted absorption components 
of Na\,\textsc{i} D, Li\,\textsc{i} $\lambda$6707 and Fe\,\textsc{ii} $\lambda$5018. Although the `centre of power' for periodic variations in these lines are 
slightly different, ($13.27$, $12.89$, $14.27$, $14.81$ and $13.57$ days respectively) the range of allowed periods at a false alarm probability of $0.01$ all 
overlap and we will assume that the same wind period, (a variance weighted average of $13.48$ days) applies to all these lines.

ii) We have also confirmed the periodic variation of the cross-correlation function constructed from spectral regions that, lacking strong wind features, 
are believed to originate from the photosphere of the disc. The detected period (around $3.64$ days in all the spectral windows investigated) is close 
to the value of $3.54$ days found by \citet{herbig03}. We find that variability is largely confined to low velocities on the blue wing of the 
cross-correlation function ($> -18.5$ km s$^{-1}$ and with a velocity centroid at $\sim -9.0$ km s$^{-1}$); a change in the characteristics of the outflow in 
FU Orionis may explain why \citet{herbig03}, with a velocity resolution of $\sim 13$ km s$^{-1}$ compared with $4$ km s$^{-1}$ in the present study, also found 
significant variation on the red wing.

In what follows we will assume that the $13.48$ period is associated with modulation of the wind. The fact that the $3.64$ day period is only 
detectable on the blue wing of the cross correlation function raises the question of whether it might not also be a wind period, even though 
the spectral regions from which the cross correlation was calculated are believed to originate in the disc photosphere. We are dissuaded 
from invoking a wind component in these lines on the grounds that in any case the period is quite different from the `wind period' described 
in i) above. We note that the two periods are {\it consistent} with being in the ratio of $1:4$ although the range of allowed periods at a false 
alarm probability of $0.01$ or below does not allow us to assert this with any confidence. Consequently, we will assume that there is no necessary 
connection between the phenomena giving rise to these two periods and that they are associated with two physically distinct regions, the wind and 
the disc photosphere. We will briefly revisit this assumption at the end of this section.

\subsection{The periodicity of the wind}

We have found clear evidence (see the upper four panels of Fig. \ref{fig:phase}) that the variations in the redshifted emission of the H$\alpha$ and 
H$\beta$ lines are in-phase with each other, with the blue shifted absorption of the Na\,\textsc{i} D and Li I $\lambda$6707 displaying an intermediate 
phase. The variation of the blueshifted absorption of H$\alpha$ is however phase lagged by about $1.8$ days.

This pattern is consistent with a picture in which the wind is modulated at its base in velocity and/or density/temperature. The first lines to respond are 
those formed at relatively low velocities close to the wind base while the blue shifted H$\alpha$ absorption (which is at considerably higher 
velocities) responds after a time equal to the flow time across the region of acceleration. This hypothesis allows us to estimate the scale of the 
region where the wind is accelerated up to $\sim 100$ km s$^{-1}$ as being the product of $100$ km s $^{-1}$ and $1.8$ days, i.e. $\sim 10^{12}$ 
cm. It should be noted however, that the phase of the Fe\,\textsc{ii} line does not fit this picture. The lack of significant periodic behaviour at still 
higher velocities could either be due to the shearing out of high density pulses in the wind beyond this region or else because the higher velocity 
absorption originates from a larger region in which a number of consecutive pulses are always simultaneously present.

\subsection{The periodicity of photospheric absorption}

A simple interpretation of periodic variability in the line profiles from a disc is in terms of an orbiting hotspot which enhances the 
flux (and hence contribution to the absorption line spectrum) at a particular velocity. The problem with this simplest model is that
it would predict that enhanced absorption should switch between the blue and red wings of the line as the hotspot alternatively approaches 
and recedes from the observer. However, we find little variability on the red wing of the line and so we need to argue that the receding hotspot 
is in some way concealed from the observer. This necessarily involves the invoking of three dimensional effects in the disc such as a warp. If 
for some reason the disc surface is distorted in the region of enhanced dissipation then the radiation may escape preferentially 
in the forward direction of the orbiting feature.

So far we have been deliberately unspecific about the nature of the hotspot or warp. We bear in mind that our confirmation of 
Herbig's period implies that this is a phenomenon that is stable over of order a thousand orbits or more and is thus inconsistent with 
a transient heating event which would be sheared out over many fewer orbital periods. \citet{clarke03} suggested that a long-lived 
hotspot would be produced if the inner disc of FU Orionis contained an embedded (proto-)hot Jupiter, as required by the model 
for the triggering of rapid-rise objects (such as FU Orionis and V1057 Cygni) suggested by \citet{clarke96} and \citet{lodato04}. 
In this case, the `hotspot' is simply the emergence of the energy liberated by accretion on to the planet as it diffuses out through 
the overlying disc. If, in addition, the disc is warped in the vicinity of the planet, (as would result if the planet and disc 
planes are not perfectly aligned) then this radiation will preferentially emerge through the region of the disc surface which offers the lowest 
optical depth to escaping photons; thus there are geometries for which the hotspot signature would be more detectable when the planet 
is approaching the observer than on the opposite phase.

In a similar vein, one could instead invoke a magnetised hotspot on the disc surface, (supported by the detection by \citet{donati05}) and again, if the axis 
of the magnetic field is misaligned with that of the disc, one expects a warped structure in the disc \citep{lai99,lai03,foucart11}. The stability of the 
observed period implies that the magnetic topology needs to remain constant on time-scales of $\sim 10$ years.

Note that it is not a necessary ingredient of the model that the hotspot co-rotates with the local disc flow, although obviously 
in the planet case, both hotspot and disc orbit at the local Keplerian velocity. If this is the case then a period of $3.6$ days corresponds 
to an orbital radius of $ 7 \times 10^{11} M^{1/3}$ cm where $M$ is the mass of the central star in solar masses. {\footnote{Note that this 
radius is factor $2-3$ larger than the best-fitting inner disc radius in FU Orionis employed by \cite{kenyon88}}} The maximum velocity shift would in 
this case be $140 M^{1/3} \rm{sin i}$ km s$^{-1} > 60 M^{1/3}$ km s$^{-1}$ where in the latter inequality we have adopted the minimum value of sin 
i quoted in the literature, \citep{kenyon88} and $M$ is the stellar mass in solar masses. Even in the case of rather a low mass star $M \sim 0.3$, 
the resulting peak projected velocity is around $40$ km s $^{-1}$. Inspection of Figs. \ref{fig:6138ccf}, \ref{fig:6320ccf} and \ref{fig:5540ccf} 
show that this is close to the maximum velocity at which there is detectable variation on the blue wing of the cross correlation function, 
although a significant period is only detectable for velocities less than $18.5$ km s$^{-1}$. The amplitude of the velocity variation is thus marginally 
consistent with a Keplerian hotspot but only if one adopts rather low values for the stellar mass and inclination of FU Orionis.

On the other hand, we have more freedom in the case of a magnetically generated hotspot since in this case the concentration of magnetic 
field lines associated with the hotspot are not necessarily co-orbital with the disc - it is now possible for the combination of low velocity 
amplitude and short period to be associated with an interaction between the field and the disc occurring well outside co-rotation. 
Note that if one identified the rotation period of the magnetic field with that of the star, then it would be rotating at 
a velocity modestly less than break-up velocity.

\subsection{The relationship between the periods}

We finally return to the issue of whether there can be any connection between the two varieties of periodic behaviour described above. The 
phase relationship between the low and high velocity wind components suggests that we are witnessing the effects of pulsed input at the 
base of the wind flow rather than the effects of rotational modulation of an intrinsically steady but non-axisymmetric wind. 
In this case we need some process to modulate the injection of momentum into the wind close to its base where it interacts with the disc. 
This process does not share the same period as the modulation of the photospheric lines. Nevertheless it is possible in principle for the warped 
structure that we have invoked to explain the modulation of the photospheric lines may also have some influence on modulating the mass uptake in 
the wind (note the similarity in spatial scales between the inferred radius of a Keplerian hotspot and the vertical height in the wind from which 
we infer that the variable wind signatures originate.) This would require that the warped structure, in addition to rigid rotation at the orbital 
period of the putative hotspot, also underwent low amplitude tilt motions normal to the disc plane. Clearly we have no detailed calculations to 
support this suggestion but merely point out that this could represent an economical solution to the `two period' problem.

\section*{Acknowledgments}
We thank Nuria Calvet, Giuseppe Lodato and Marina Rominova for helpful discussions regarding physical interpretations of the observed cyclic variability, as 
well as the referee for the helpful comments.
SLP thanks the Science and Technology Facilities Council (STFC) for the award of a PhD studentship and acknowledges funding through \textsc{constellation}, 
an European Commission FP6 Marie Curie Research Training Network, and Churchill College, Cambridge for travel grants. SLP would also like to thank James 
Owen for invaluable technical support.

\bibliographystyle{mn2e}
\bibliography{references}


\appendix

\label{lastpage}

\end{document}